\documentclass[review]{elsarticle}

\usepackage[utf8]{inputenc}
\usepackage[T1]{fontenc}

\usepackage{amsmath, bm}
\usepackage{amssymb}
\usepackage{graphicx,psfrag,epsf}
\usepackage{enumerate}
\usepackage{natbib}

\usepackage{graphicx}

\usepackage{hyperref}
\usepackage{float}
\usepackage{subfig}
\usepackage{wrapfig}
\usepackage{rotating}
\usepackage{lscape}
\usepackage[T1]{fontenc}

\usepackage{amssymb}
\usepackage{tabularx}
\usepackage[dvips]{epsfig}

\date{}

\usepackage{dsfont}

\usepackage{multirow}
\usepackage{xcolor}
\usepackage[normalem]{ulem}

\newcommand*{\rom}[1]{\expandafter\@slowromancap\romannumeral #1@}

    \usepackage{hyperref} 
\hypersetup{colorlinks=true, linkcolor=blue,
  citecolor=blue,urlcolor=blue} 

\DeclareMathAlphabet{\mathcal}{OMS}{cmsy}{m}{n}
\SetMathAlphabet{\mathcal}{bold}{OMS}{cmsy}{b}{n}

\newcommand{\vertiii}[1]{{\left\vert\kern-0.25ex\left\vert #1 
    \right\vert\kern-0.25ex\right\vert_F}}

\renewcommand{\vec}[1]{\;\textsl{vec}\!\left(\bm{#1}\right)}

\usepackage{color} 
\usepackage[normalem]{ulem} 

    \newcommand{\ta}[1]{#1} 
    \newcommand{\tO}[1]{#1} 
    \newcommand{\tr}[1]{#1} 
    \newcommand{\td}[1]{} 

\newcommand{\rtild}{\widetilde{R}_k}

\newcommand{\sigtau}[1]{\left(\sigma_\beta\tau_{#1}\right)}
\newcommand{\sqrtcov}{\left({\bm{\Sigma}_Y}^{-1/2} \otimes \bm{I}_n\right)}
\newcommand{\Xk}[1]{\left(\bm{\Sigma}_Y^{-1/2} \otimes \bm{X}_{#1}\right)}
\newcommand{\Xkp}[1]{\left(\bm{\Sigma}_Y^{-1/2} \otimes \bm{X}_{#1}'\right)}

\newcommand{\covXkkX}[1]{\left(\bm{\Sigma}_Y^{-1} \otimes \bm{X}_{#1}\bm{X}_{#1}'\right)}
\newcommand{\covkXXk}[1]{\left(\bm{\Sigma}_Y^{-1} \otimes \bm{X}_{#1}'\bm{X}_{#1}\right)}
\newcommand{\fullcov}{\left({\bm{\Sigma}_Y}^{-1} \otimes \bm{I}_n\right)}

\newtheorem{thm}{Theorem}[section]
\newtheorem{lemma}{Lemma}

\makeatletter
\renewcommand*{\@fnsymbol}[1]{\ifcase#1\or*\else\fi}
\makeatother

\journal{ }

\begin{document}

\begin{frontmatter}

\title{An Approach of Bayesian Variable Selection for Ultrahigh Dimensional Multivariate Regression}

\author{Xiaotian Dai}
\address{Department of Mathematical Sciences, SUNY Binghamton University, Vestal, NY 13850.}

\author{Guifang Fu\footnote{Corresponding author: Guifang Fu. Department of Mathematical Sciences, SUNY Binghamton University, Vestal, NY 13850;  Email: gfu@binghamton.edu; Tel: +1(607) 777-3550.}}
\address{Department of Mathematical Sciences, SUNY Binghamton University, Vestal, NY 13850.}

\author{Randall Reese}
\address{Idaho National Laboratory, Idaho Falls, ID 83402.}

\author{Shaofei Zhao}
\address{Department of Mathematical Sciences, SUNY Binghamton University, Vestal, NY 13850.}

\author{Zuofeng Shang}
\address{Department of Mathematical Sciences, New Jersey Institute of Technology, Newark, NJ 07102.}

\begin{abstract}
In many practices, scientists are particularly interested in detecting which of the predictors are truly associated with a multivariate response. It is more accurate to model multiple responses as one vector rather than separating each component one by one. This is particularly true for complex traits having multiple correlated components. \tr{A Bayesian multivariate variable selection (BMVS) approach is proposed} to select important predictors influencing the multivariate response from a candidate pool with ultrahigh dimension. By applying the sample-size-dependent spike and slab priors, \tr{the BMVS approach satisfies} the strong selection consistency property under certain conditions, which represents the advantages of BMVS over other existing Bayesian multivariate regression-based approaches. \tr{The proposed approach considers} the covariance structure of multiple responses without assuming independence, and integrates the estimation of covariance related parameters together with all regression parameters into one framework through a fast updating MCMC procedure. \tr{It is demonstrated} through simulations that the BMVS approach outperforms some other relevant frequentist and Bayesian approaches. The proposed BMVS approach possesses a flexibility of wide applications, including genome-wide association studies with multiple correlated phenotypes and a large scale of genetic variants and/or environmental variables, as demonstrated in the real data analyses section. The computer code and test data of the proposed method is available as an R package.
\end{abstract}


\end{frontmatter}


\section{Introduction}
Modeling the association between a multivariate response and a set of predictors has been attracting considerable attention in various disciplines, including ecology, geology, psychology, genetics, among others. For example, De'Ath et al. \cite{de2002multivariate} modeled the relationship between multi-species and environmental factors. Tsitsika et al. \cite{tsitsika2009internet} selected important predictors associated with multiple internet addiction characteristics for adolescents. Influential single-nucleotide polymorphisms (SNPs) affecting multiple biological traits were detected through genome-wide association studies (GWAS) \cite{yang2010analyze, o2012multiphen, bottolo2013, zhou2014efficient, liquet2016}.   When the number of predictors is much larger than the number of observations  ($p > n$) \cite{o2012multiphen, scott2012large}, overfitting and low prediction accuracy may become problematic. Therefore, variable selection is crucial for high dimensional data analyses. 

As a well-known frequentist method, penalized regression has been widely used for variable selection purposes in high-dimensional data analyses \cite{li2011bayesian, rothman2010sparse}. These include LASSO \cite{tibshirani1996}), adaptive LASSO \cite{zou2006adaptive}, and Elastic net \cite{zou2005regularization}, to name a few. Despite LASSO's popularity in shrinking unimportant predictors to zero, its theoretical properties are controversial. Some research claims that the selection consistency of LASSO cannot be established and its performance for high-dimensional predictors is not well explored \cite{fan2001variable, zou2006adaptive, narisetty2014bayesian}. Breiman and Friedman \cite{breiman1997predicting} proposed the Curds and Whey algorithm to enhance the prediction accuracy of multivariate regression by taking advantage of the correlations between response variables. However, their algorithm fitted a full model using $\bm{\ell}_2$ penalty, which cannot shrink unimportant predictors to produce a sparse model \cite{fan2001variable}. She and Chen \cite{she2017} proposed using reduced-rank regression to handle high-dimensional multivariate data analysis, and set a penalty constraint not only on the regression coefficients but also on the rank of the coefficient matrix. The disadvantages of the reduced rank approach included its sensitiveness on the rank constraint and outliers of the data for low-rank coefficient matrix estimation \cite{she2017}. In addition, the theoretical properties of reduced-rank regression investigated by She and Chen \cite{she2017} focused on non-asymptotic robust analysis and prediction accuracy improvements, instead of the asymptotic selection consistency. 

Bayesian variable selection approaches have been well-established in univariate regression settings where there is only one response variable. Consider a univariate regression model here: $\boldsymbol{Y}_{n \times 1} = \boldsymbol{X}_{n \times p}\boldsymbol{\beta}_{p \times 1} + \boldsymbol{\epsilon}_{n \times 1}$. Bayesian methods usually introduce a latent variable $Z_i$ for each predictor, in which $Z_i = 0$ or 1, $i=1,\ldots,p$ indicates whether or not the $i^{th}$ predictor should be included into the final model. Hence a binary vector $\textbf{Z}$ of length $p$ can indicate a particular model. There have been many efforts devoted to develop ``objective'' priors for selecting the optimal model, such as the Jeffreys prior, Zellner's $g$-prior, and their related model selection literature \cite{zellner1980, zellner1984, laud1995predictive, kass1995reference, berger1996intrinsic, moreno1998intrinsic, de1999methods, perez2002expected, bayarri2008generalization, liang2008mixtures, cui2008empirical, maruyama2010gbf, maruyama2010robust, johnson2012bayesian, cano2013integral, bove2015objective, fouskakis2016power}. The word ``objective'' indicates that these priors were not ``subjective'' and applicable to various model selection problems \cite{bayarri2012}.

As the advent of new data collection technologies, Bayesian regularized regression, such as the Bayesian lasso \cite{park2008bayesian, li2011bayesian} and the spike and slab priors, was investigated to cope with the challenges of high dimensionality. The connections between regularized regressions and the spike and slab priors have been studied in \cite{narisetty2014bayesian}. Mitchell and Beauchamp \cite{mitchell1988bayesian} proposed the use of spike and slab priors on regression coefficients $\beta$'s. When $Z_i = 0$, the $i^{th}$ predictor is considered inactive and $\beta_i$ will have a prior distribution with concentrated probability mass around zero, which is referred to as the spike prior. When $Z_i = 1$, the $i^{th}$ predictor is considered active in the model and $\beta_i$ will have a prior distribution with diffusing probability density, which is referred to as the slab prior. George and McCulloch \cite{george1993variable} proposed a stochastic search variable selection (SSVS) approach to select ``promising'' subsets of predictors. Their framework used a normal distribution with zero mean and a small fixed variance as the spike prior, and another normal distribution with zero mean and a large fixed variance as the slab prior. Most recently, Narisetty and He \cite{narisetty2014bayesian} questioned the theoretical selection consistency property for the approaches using fixed hyperparameters for spike and slab priors, and showed that SSVS did not guarantee selection consistency. As the first work changing the fixed variances of the spike and slab priors into sample-size-dependent shrinking and diffusing priors, Narisetty and He \cite{narisetty2014bayesian} proved that the criteria for strong selection consistency were satisfied in univariate regression \cite{narisetty2014bayesian}. 

In terms of how many variables should be selected, traditional Bayesian literature employed a latent vector $\boldsymbol{Z}_{p \times 1}$ to indicate the inclusions and exclusions of all the $p$ predictors, and then the optimal model was determined by finding the latent vector yielding the highest model posterior probability \cite{zellner1980, laud1995predictive, berger1996intrinsic, moreno1998intrinsic, perez2002expected, liang2008mixtures, cui2008empirical, maruyama2010gbf, maruyama2010robust, shang2011consistency, fouskakis2016power}. Barbieri and Berger \cite{barbieri2004} suggested that an optimal model in the Bayesian variable selection was often the ``median probability model'', which was defined to be the model consisting of the predictors whose overall posterior inclusion probability is at least 1/2. The overall posterior inclusion probability is the sum of the posterior probabilities of the models which contain that predictor. In a recent study of Univariate Bayesian variable selection, Narisetty and He \cite{narisetty2014bayesian} proposed to use marginal posterior probability of $\{Z_i=1\}$ to select the optimal model and showed that the selection consistency can be guaranteed using sample-size-dependent shrinking and diffusing priors. This is a major improvement because the use of marginal posterior probability is definitely computationally advantageous compared to computing the model posterior probabilities among the entire model space.

Unlike the aforementioned univariate variable selection approaches, variable selection approaches for multivariate regressions are undeveloped. Brown et al. \cite{brown1998multivariate} developed a Bayesian method for multivariate variable selection, but their approach inherited aforementioned limitation of using model posterior probabilities. Therefore, it was not computationally feasible when $p$ is greater than 20 since it involves $2^p$ choices \cite{brown1998multivariate}. A more recent development on Bayesian multivariate regression-based variable selection approach was the GUESS algorithm proposed by Bottolo et al. \cite{bottolo2013} and the R package implementing the GUESS algorithm \cite{liquet2016}. Compared to the aforementioned Bayesian univariate regression-based approaches, GUESS was able to model multiple correlated responses and their covariance structure. However, similar to SSVS, the GUESS algorithm used the fixed hyperparameters for spike and slab priors, and hence did not have the theoretical guarantees about the selection consistency. Moreover, GUESS also employed model posterior probabilities to search the optimal model from entire model space. To overcome the computational cost of model posterior probabilities, Bottolo et al. \cite{bottolo2013} adopted Graphical Processor Unit (GPU) technologies, specifically using NVIDIA's Complete Unified Device Architecture (CUDA). \tr{Most recently, Ning et al. \cite{ning2020bayesian} proposed a novel Bayesian method to estimate regression coefficients in high-dimensional multivariate regression using a product of independent spike and slab priors. These priors accommodate sparsity at the group level, as some predictors can be naturally clustered into groups.} 

In this article, we propose a novel Bayesian multivariate variable selection (BMVS) approach to detect truly influential predictors that are associated with a multivariate response from an ultrahigh dimensional candidate pool (i.e., $p \gg n$ or $p = e^{o[n]}$). We consider sample-size-dependent shrinking and diffusing priors instead of fixed variance of the spike and slab priors. After modeling a complex covariance matrix without assuming independence of multiple response variables, we are still able to establish the appealing strong selection consistency for the BMVS approach under certain conditions. All unknown parameters, including the covariance-related and regression-related coefficients, are estimated through a fast-updating Markov chain Monte Carlo (MCMC) algorithm. We demonstrate that the proposed BMVS method performs well through multiple empirical simulation studies. We compare the BMVS model to a few other approaches that have been widely used in multivariate regression literature, including not only the popular frequentist methods such as canonical correlation analysis and multivariate LASSO/Elastic net, but also the Bayesian multivariate regression-based GUESS algorithm \cite{bottolo2013}. The BMVS method outperforms those relevant approaches by achieving very promising false discovery rates and power in all simulation designs. To further demonstrate its accuracy in variable selection, we apply the proposed BMVS approach to two real data analyses in genome-wide association studies. The R package will be published on the Comprehensive R Archive Network (CRAN).

\section{Methodology}
\subsection{Prior distributions and hyperparameters}
Consider the multivariate regression model:
\begin{equation}
\boldsymbol{Y} = \boldsymbol{X}\boldsymbol{\beta} + \boldsymbol{E},
\end{equation}
where $\boldsymbol{Y} \in \mathbb{R} ^ {n \times q}$ \ta{is the response matrix}, $\boldsymbol{X} \in \mathbb{R} ^ {n \times p}$ \ta{is the predictor matrix}, $\boldsymbol{\beta} \in \mathbb{R} ^ {p \times q}$ \ta{is the parameter coefficient matrix}, and $\boldsymbol{E} \in \mathbb{R} ^ {n \times q}$ is the error matrix with each row following an independent and identical distribution $N_q(\boldsymbol{0}, \boldsymbol{\Sigma}_{Y})$. \ta{Here $p$ denotes the number of predictors, $q$ denotes the number of response variables, and $n$ denotes the number of observations. Let} $\boldsymbol{Y}_m$ and $\boldsymbol{X}_m$ denote the $m^{th}$ row vector in $\boldsymbol{Y}$ and $\boldsymbol{X}$, respectively, where $m = 1,...,n$. \ta{The $i^{th}$ row vector of $\boldsymbol{\beta}$, $\boldsymbol{\beta}_i$, having a $1 \times q$ dimension, corresponds to the coefficient vector of the $i^{th}$ predictor, where $i = 1,...,p$.} For selection purpose, we assume that \ta{the true model is sparse (i.e., active/non-active ratio is small). Both the regression coefficients $\boldsymbol{\beta}$  and the covariance matrix $\boldsymbol{\Sigma}_{Y}$  are unknowns that need to be estimated}.

For each of the predictors, we introduce a latent binary variable $Z_i$ to decide whether or not it should be included in the model. \ta{An active predictor is indicated by $Z_i = 1$ and a non-active predictor by $Z_i = 0$, with distribution $\mathbb{P}(Z_i = 1) = 1 - \mathbb{P}(Z_i = 0) = \phi$. Intuitively, the priors of $\phi$ should be given by the influential/non-influential ratio of the candidate predictor pool, {which nevertheless is unknown in real datasets. We will empirically verify through simulations that the choice of priors for $\phi$ has trivial impacts on the final result}.

The priors of the regression coefficients $\boldsymbol{\beta}$ can be given as}
\begin{equation}\label{model}
\begin{gathered}
\bm{Y} | (\bm{X}, \bm{\beta}, \bm{\Sigma}_Y, \sigma_\beta^2) \sim N_q(\bm{X}\bm{\beta}, \bm{\Sigma}_Y),\\
\boldsymbol{\beta}_i | (\sigma_{\beta}^2, Z_i = 1) \sim N_q(\boldsymbol{0}, \sigma_{\beta}^2 \tau_1^2 \boldsymbol{I}_q),\\
\boldsymbol{\beta}_i | (\sigma_{\beta}^2, Z_i = 0) \sim N_q(\boldsymbol{0}, \sigma_{\beta}^2 \tau_0^2 \boldsymbol{I}_q),\\
\text{and } \sigma_{\beta}^2 \sim IG(\alpha_1, \alpha_2),
\end{gathered}
\end{equation}
where $\boldsymbol{I}_q$ is a $q$ by $q$ identity matrix. $\sigma_{\beta}^2$ is a scalar parameter following an inverse gamma distribution with a shape parameter $\alpha_1$ and a scale parameter $\alpha_2$, as the inverse gamma distribution is conditionally conjugate and the priors are essentially non-informative when $\alpha_1$ and $\alpha_2$ are reasonably small \cite{gelman2006prior}.

The sample-size-dependent parameters $\tau_1^2$ and $\tau_0^2$ control the priors of $\boldsymbol{\beta}_i$ to make it either a slab prior or a spike prior. If $Z_i = 0$, $\boldsymbol{\beta}_i$ will have a spike prior distribution with concentrated probability mass around zero. If $Z_i = 1$, $\boldsymbol{\beta}_i$ will have a slab prior distribution with diffusing probability density. Inspired by Narisetty and He \cite{narisetty2014bayesian}, we design the prior hyperparameters of $ \tau_0^2$ and $\tau_1^2 $ as
\begin{equation}
\begin{gathered}
\tau_0^2 = \frac{\hat{\overline{\sigma}}_Y^2}{10n},\\
\text{and } \tau_1^2 = \hat{\overline{\sigma}}_Y^2\text{max}\bigg(\frac{(pq)^{2.1}}{100n},\text{log}n \bigg),
\end{gathered}
\end{equation}
where $\hat{\overline{\sigma}}_Y^2$ is the average sample variances of multiple response variables $\hat{\overline{\sigma}}_Y^2 = \sum_{k=1}^q {{\hat{\sigma}_{Y_k}}}^2 / q$. These choices of $\tau_1^2$ and $\tau_0^2$ asymptotically satisfy the conditions that $\tau_1^2 \rightarrow \infty$ and $\tau_0^2 \rightarrow 0$ as $n \rightarrow \infty$.

A natural choice for the prior distribution of covariance matrix $\boldsymbol{\Sigma}_{Y}$ would be a probability distribution defined on positive-definite matrices with the conjugate property. According to Dawid \cite{dawid1981some}, directly modeling the matrix variate has the advantage of preserving the matrix structures without breaking the matrices down into multiple row or column vectors that dramatically increase complexity and computation costs. As such, we choose the inverse Wishart distribution:
\begin{equation}
\boldsymbol{\Sigma}_{Y} \sim IW(\nu, \Lambda),
\end{equation}
where $\nu$ is the degree of freedom, and $\Lambda$ is a positive-definite scale matrix. The default values for these prior hyperparameters are $\nu = q + 1$ and $\Lambda = \boldsymbol{I}_q$ \cite{leonard1992bayesian}. Due to the conjugacy of the inverse Wishart distribution, $\boldsymbol{\Sigma}_{Y}$ can be conveniently estimated together with all the other unknown parameters.

\subsection{Posterior distributions}
Assuming that $Z_i$'s and $\sigma_{\beta}^2$ are conditionally independent of $\boldsymbol{\Sigma}_{Y}$, the joint posterior distribution containing all unknown parameters can be derived as:
\begin{small}
\begin{align*}
f&(\boldsymbol{\beta}, \boldsymbol{Z}, \sigma_{\beta}^2, \boldsymbol{\Sigma}_{Y} | \boldsymbol{Y}) \\
& \propto f(\boldsymbol{Y} | \boldsymbol{\beta}, Z, \sigma_{\beta}^2, \boldsymbol{\Sigma}_{Y}) \times \prod_{i=1}^{p} f(\boldsymbol{\beta}_i) \times \prod_{i=1}^{p} f(Z_i) \times f(\sigma_{\beta}^2) \times f(\boldsymbol{\Sigma}_{Y}) \\
& \propto \bigg(\frac{1}{\sqrt{|\Sigma_Y|}} \bigg)^n \text{exp}\bigg(-\frac{1}{2} \sum_{m=1}^{n}(\boldsymbol{Y}_m - \boldsymbol{X}_m\boldsymbol{\beta})\boldsymbol{\Sigma}_{Y}^{-1}(\boldsymbol{Y}_m - \boldsymbol{X}_m\boldsymbol{\beta})' \bigg) \\
& \quad \times \prod_{i=1}^{p} \bigg((1 - \phi)\frac{1}{\sqrt{|\sigma_{\beta}^2 \tau_0^2 \boldsymbol{I}_q|}} \text{exp}(-\frac{1}{2}\boldsymbol{\beta}_i(\sigma_{\beta}^2 \tau_0^2 \boldsymbol{I}_q)^{-1}\boldsymbol{\beta}_i')  \bigg)^{1-Z_i} \\
&  \quad \times \prod_{i=1}^{p} \bigg(\phi\frac{1}{\sqrt{|\sigma_{\beta}^2 \tau_1^2 \boldsymbol{I}_q|}} \text{exp}(-\frac{1}{2}\boldsymbol{\beta}_i(\sigma_{\beta}^2 \tau_1^2 \boldsymbol{I}_q)^{-1}\boldsymbol{\beta}_i')  \bigg)^{Z_i} \\
& \quad \times \sigma_{\beta}^{-2(\alpha_1+1)}\text{exp}\bigg(-\frac{\alpha_2}{\sigma_{\beta}^2} \bigg) \times |\Sigma_Y|^{-(\nu+q+1)/2} \text{exp}\bigg(tr(\Lambda\Sigma_Y^{-1})\bigg).
\end{align*}
\end{small}

The \tr{full conditional posterior} distribution of $\boldsymbol{\beta}_i$, the coefficients of the $i^{th}$ predictor, is
\begin{small}
\begin{align*}
f&(\boldsymbol{\beta}_i | Z_i, \sigma_{\beta}^2, \boldsymbol{\Sigma}_{Y}, \boldsymbol{Y}) \\
& \propto f(\boldsymbol{Y} | \boldsymbol{\beta}_i, Z_i, \sigma_{\beta}^2, \boldsymbol{\Sigma}_{Y}) \times f(\boldsymbol{\beta}_i) \\
& \propto \text{exp}\bigg(-\frac{1}{2} \sum_{m=1}^{n}(\boldsymbol{Y}_m - \boldsymbol{\mu}_{m(-\boldsymbol{\beta}_i)} - \boldsymbol{X}_{im}\boldsymbol{\beta}_i) \boldsymbol{\Sigma}_{Y}^{-1} \\
& \quad \quad \quad \quad \quad \quad \times (\boldsymbol{Y}_m - \boldsymbol{\mu}_{m(-\boldsymbol{\beta}_i)} - \boldsymbol{X}_{im}\boldsymbol{\beta}_i)' \\ 
& \quad \quad \quad \quad \quad \quad \quad \quad \quad -\frac{1}{2} \boldsymbol{\beta}_i (\sigma_{\beta}^2 \tau_i^2 \boldsymbol{I}_q)^{-1} \boldsymbol{\beta}_i' \bigg) \\
& \propto \text{exp}\bigg(\boldsymbol{\beta}_i((\sigma_{\beta}^2 \tau_i^2 \boldsymbol{I}_q)^{-1} + \sum_{m=1}^{n} X_{mi}^2 \boldsymbol{\Sigma}_{Y}^{-1})\boldsymbol{\beta}_i'\\
&\quad \quad \quad\quad \quad \quad\quad \quad \quad - 2\sum_{m=1}^{n}(\boldsymbol{Y}_m - \boldsymbol{\mu}_{m(-\boldsymbol{\beta}_i)})\boldsymbol{\Sigma}_{Y}^{-1}(X_{mi}\boldsymbol{\beta}_i) \bigg) \\
& \propto N_q(\mu_{\boldsymbol{\beta}_i}, \Sigma_{\boldsymbol{\beta}_i}),
\end{align*}
\end{small}
where $\boldsymbol{\mu}_{(-\boldsymbol{\beta}_i)} = \boldsymbol{X}\boldsymbol{\beta} - \boldsymbol{X}_i\boldsymbol{\beta}_i$. We also have
\begin{small}
\begin{align*}
\mu_{\boldsymbol{\beta}_i} = &\bigg((\sigma_{\beta}^2 \tau_i^2 \boldsymbol{I}_q)^{-1} + \sum_{m=1}^{n} X_{mi}^2 \boldsymbol{\Sigma}_{Y}^{-1} \bigg)^{-1} \\ 
&\times \bigg(\sum_{m=1}^{n}(\boldsymbol{Y}_m - \boldsymbol{\mu}_{m(-\boldsymbol{\beta}_i)})\boldsymbol{\Sigma}_{Y}^{-1}X_{mi} \bigg)', \\
\Sigma_{\boldsymbol{\beta}_i} = &\bigg((\sigma_{\beta}^2 \tau_i^2 \boldsymbol{I}_q)^{-1} + \sum_{m=1}^{n} X_{mi}^2 \boldsymbol{\Sigma}_{Y}^{-1} \bigg)^{-1},
\end{align*}
\end{small}
where $X_{mi}$ is the $m^{th}$ observation of the $i^{th}$ predictor.

The \tr{conditional posterior} distribution of $\sigma_{\beta}^2$ is
\begin{small}
\begin{align*}
f&(\sigma_{\beta}^2 | \boldsymbol{\beta}, \boldsymbol{Z}, \boldsymbol{Y}) \\
& \propto f(\boldsymbol{Y} | \boldsymbol{\beta}_i, Z_i, \sigma_{\beta}^2, \boldsymbol{\Sigma}_{Y}) \times f(\sigma_{\beta}^2) \\
& \propto \sigma_{\beta}^{-2(\alpha_1+1)}\text{exp}\bigg(-\frac{\alpha_2}{\sigma_{\beta}^2} \bigg) \\
& \quad \times \prod_{i=1}^{p} \bigg(\frac{1}{\sqrt{|\sigma_{\beta}^2 \tau_0^2 \boldsymbol{I}_q|}} \text{exp}(-\frac{1}{2}\boldsymbol{\beta}_i(\sigma_{\beta}^2 \tau_0^2 \boldsymbol{I}_q)^{-1}\boldsymbol{\beta}_i')  \bigg)^{1-Z_i}\\
& \quad \times \prod_{i=1}^{p} \bigg(\frac{1}{\sqrt{|\sigma_{\beta}^2 \tau_1^2 \boldsymbol{I}_q|}} \text{exp}(-\frac{1}{2}\boldsymbol{\beta}_i(\sigma_{\beta}^2 \tau_1^2 \boldsymbol{I}_q)^{-1}\boldsymbol{\beta}_i')  \bigg)^{Z_i} \\
& \propto IG\bigg(\alpha_1 + pq, \alpha_2 + \sum_{i=1}^{p} \boldsymbol{\beta}_i(\tau_{Z_i}^2 \boldsymbol{I}_q)^{-1}\boldsymbol{\beta}_i' \bigg).
\end{align*}
\end{small}

\tr{
The conditional posterior distribution of $Z_i$ is
\begin{small}
\begin{align*}
f&(Z_i | \boldsymbol{\beta}_i, \sigma_{\beta}^2) \\
& \propto \bigg((1 - \phi)\frac{1}{\sqrt{|\sigma_{\beta}^2 \tau_0^2 \boldsymbol{I}_q|}} \text{exp}(-\frac{1}{2}\boldsymbol{\beta}_i(\sigma_{\beta}^2 \tau_0^2 \boldsymbol{I}_q)^{-1}\boldsymbol{\beta}_i')  \bigg)^{1-Z_i} \\
&  \quad \times \bigg(\phi\frac{1}{\sqrt{|\sigma_{\beta}^2 \tau_1^2 \boldsymbol{I}_q|}} \text{exp}(-\frac{1}{2}\boldsymbol{\beta}_i(\sigma_{\beta}^2 \tau_1^2 \boldsymbol{I}_q)^{-1}\boldsymbol{\beta}_i')  \bigg)^{Z_i},
\end{align*}
\end{small}

and $\mathbb{P}(Z_i = 1 | \boldsymbol{\beta}_i, \sigma_{\beta}^2) + \mathbb{P}(Z_i = 0 | \boldsymbol{\beta}_i, \sigma_{\beta}^2) = 1$.

Therefore, the \tr{conditional} posterior probability of $Z_i$ is
\begin{small}
\begin{align*}
\mathbb{P}(Z_i = 1 | \boldsymbol{\beta}_i, \sigma_{\beta}^2) = \frac{\phi \theta_q (\boldsymbol{\beta}_i; \boldsymbol{0}, \sigma_{\beta}^2 \tau_1^2 \boldsymbol{I}_q)}{\phi \theta_q (\boldsymbol{\beta}_i; \boldsymbol{0}, \sigma_{\beta}^2 \tau_1^2 \boldsymbol{I}_q) + (1 - \phi) \theta_q (\boldsymbol{\beta}_i; \boldsymbol{0}, \sigma_{\beta}^2 \tau_0^2 \boldsymbol{I}_q)},
\end{align*}
\end{small}
where $\theta_q (\boldsymbol{\beta}_i; .)$ is the value of a $q$-dimensional multivariate normal distribution evaluated at $\boldsymbol{\beta}_i$.
}

Finally, the \tr{conditional posterior} distribution of $\boldsymbol{\Sigma}_{Y}$ is
\begin{small}
\begin{align*}
f&(\boldsymbol{\Sigma}_{Y} | \boldsymbol{\beta}, \boldsymbol{Y}) \\
& \propto f(\boldsymbol{Y} | \boldsymbol{\Sigma}_{Y}, \boldsymbol{\beta}_i, Z_i, \sigma_{\beta}^2) \times f(\boldsymbol{\Sigma}_{Y}) \\
& \propto \bigg(\frac{1}{\sqrt{|\Sigma_Y|}} \bigg)^n \text{exp}\bigg(-\frac{1}{2} \sum_{m=1}^{n}(\boldsymbol{Y}_m - \boldsymbol{X}_m\boldsymbol{\beta})\boldsymbol{\Sigma}_{Y}^{-1}(\boldsymbol{Y}_m - \boldsymbol{X}_m\boldsymbol{\beta})' \bigg) \\
& \quad \times |\Sigma_Y|^{-(\nu+q+1)/2} \text{exp}\bigg(tr(\Lambda\Sigma_Y^{-1})\bigg) \\
& \propto IW(n + \nu, \Lambda + \sum_{m=1}^{n} (\boldsymbol{Y}_m - \boldsymbol{X}_m\boldsymbol{\beta})(\boldsymbol{Y}_m - \boldsymbol{X}_m\boldsymbol{\beta})').
\end{align*}
\end{small}

\subsection{Estimation}
Since all conditional posterior distributions have standard forms, we use Gibbs samplers to simulate and estimate them. The selection of the final model is based on the estimated marginal posterior probabilities of $Z_i$'s, a higher of which indicates stronger associations between the corresponding predictor and the multivariate response. Therefore, we rank predictors by their estimated posterior probabilities $\hat{\mathbb{P}}(Z_i = 1 | \boldsymbol{\beta}_i, \sigma_{\beta}^2)$. To determine how many predictors to keep and to identify the optimal model, \tr{we use the multivariate corrected Akaike information criterion (AICc) proposed by Bedrick and Tsai \cite{bedrick1994model}:

\begin{center}
AICc = $n log|\hat{\boldsymbol{\Sigma}}_{Y}| + dq(n + p)$,
\end{center}
where $d = n / \{n - (p + q + 1)\}$.} The multivariate AICc considers the correlation between multiple response variables and puts a higher penalty on model size than the conventional AIC, and hence is more suitable for high-dimensional data with sparse structures; see \cite{bedrick1994model}. In addition, it works better than many other univariate-based model selection criteria that treat multiple responses separately \cite{chen2012sparse}.

\section{Simulation}
We examine the performance of the BMVS method using several simulation designs, and also compare BMVS's performance relative to many other relevant multivariate variable selection methods. We vary both the number of observations and the number of predictors. All the simulation results are calculated based on 100 replications. \tr{In each replicate, we perform 1000 burn-in iterations followed by 5000 update iterations to estimate unknown parameters.}

We perform three different simulation settings based on the standard statistical multivariate regression model described in Equation (1). The predictors in these settings are continuous variables.
\begin{itemize}
\item \emph{Setting 1}: Generate each predictor independently from a standard normal distribution: $X_{mi} \sim N(0, 1)$. Set the first ten predictors to be influential and generate their corresponding coefficients from a uniform distribution $\beta_{ij} \sim \text{Uniform}(1, 3)$ (yielding moderate effects), where $1 \leq i \leq 10$ and $1 \leq j \leq q$. Set the coefficients of all other non-causative predictors to be zero.  Let the $kj^{th}$ component of $\boldsymbol{\Sigma}_{Y}$ be $\sigma_{kj} = 0.5^{|j-k|}$, where $1 \leq k \leq q,$ \tO{and} $1 \leq j \leq q$. Finally, connect $\boldsymbol{Y}$ with the truly influential predictors using model (1). Set the number of response variables to be $q = 5$.
\item \emph{Setting 2}: Maintain all other settings from Setting 1, but only increase the difficulty level by setting the number of response variables to be $q = 30$.
\item \emph{Setting 3}: Increase the difficulty level of Setting 1 from two changes: 1) introduce correlations among predictors (instead of the independence outlined in Setting 1). The predictors are generated from a multivariate normal distribution: $\boldsymbol{X}_{m} \sim N_p(\boldsymbol{0}, \boldsymbol{\Sigma}_{X})$. The $kj^{th}$ component of $\boldsymbol{\Sigma}_{X}$ is $\sigma_{kj} = 0.5^{|j-k|}$, where $1 \leq k \leq p$ \tO{and} $1 \leq j \leq p$. \ta{2) Generate the coefficients of influential predictors} from a uniform distribution $\beta_{ij} \sim \text{Uniform}(0.5, 0.8)$, which results in very weak signals that are not easily differentiated from the non-influential candidates. Maintain all other settings from Setting 1.
\item \emph{Setting 4}: We increase the number of influential predictors to twenty (i.e., the signal-to-noise ratio is also increased accordingly), which doubles the amount used in previous three settings. In addition, a random covariance matrix is employed. Let the $kj^{th}$ component of $\boldsymbol{\Sigma}_{Y}$ to be $\sigma_{kj} = \rho^{|j-k|}$, where $\rho \sim \text{Uniform}(0.2, 0.8)$. Maintain all other settings from Setting 1.
\end{itemize}

We compare the BMVS method with three other variable selection methods: 1) canonical correlation analysis (CCA) coupled with Wilk's Lambda test to assess the significance of each predictor; 2) multivariate LASSO (M-LASSO); 3) multivariate Elastic Net (M-EN); and 4) the GUESS algorithm. We implement M-LASSO and M-EN using the R package \emph{glmnet} \cite{glmnet}, CCA using the R package \emph{CCA} \cite{gonzalez2008cca}, and the GUESS algorithm using the R package \emph{R2GUESS} \cite{liquet2016}. To make the comparisons fair, we use the multivariate AICc to determine the model size for M-LASSO, M-EN, and BMVS. The \emph{R2GUESS} package comes with its own tuning procedure to choose the optimal model by ranking the model Posterior probabilities (marginal likelihood $\times$ prior probabilities) of all choices across the model space \cite{bottolo2013}.

The results of the four simulation settings are summarized in Tables \ref{tab1.1}-\ref{tab1.4}, respectively. The true model is denoted by $t$, and the selected model is denoted by $\hat{t}$. We used five criteria to assess the performance of these approaches: the average of posterior probabilities of all influential and non-influential predictors ($mpp_1$ and $mpp_0$, respectively), the power that the exact true model is selected ($P(\hat{t} = t)$), the power that the true model is contained in the selected model ($P(\hat{t} \supset t)$), and the false discovery rate (FDR). 

Despite the fact that all approaches have essentially equal high power in achieving $\{\hat{t} \supset t\}$, \tr{the advantage of the BMVS methods is that it maintains high power in selecting the exact true model $\{\hat{t} = t\}$ while also attaining low false discovery rates.} In particular, when the M-EN and M-LASSO approaches fail to pick the exact model in any of the simulation replicate (indicated by $\mathbb{P}(\hat{t} = t)$ = 0), the BMVS method achieves astonishingly high power of nearly 100\% (see Tables \ref{tab1.1} \& \ref{tab1.2}). 
Note that it is much harder to achieve $\{\hat{t} = t\}$ than $\{\hat{t} \supset t\}$, and the difference between $\{\hat{t} = t\}$ and $\{\hat{t} \supset t\}$ is equivalent to false discoveries. It seems that the M-EN and M-LASSO approaches aggressively expand their model sizes while overfitting, scoring FDRs above 95\%, to guarantee that the true model is contained in their selected sets. The BMVS approach, on the other hand, attains low FDRs (see Tables \ref{tab1.1} - \ref{tab1.4}). The CCA approach also has a low FDR, but its power in $\mathbb{P}(\hat{t} = t)$ is only about one third of that of the BMVS approach. {\color{black}The average running times of the five methods compared (BMVS, CCA, M-LASSO, M-EN, and R2GUESS) in Setting 4 are 223s, 0.32s, 2.67s, 0.25s, and 35s respectively. The running times are evaluated on a Mac with 2.2GHz Intel Core i7 CPU and 16GB DDR4. As mentioned before, GUESS is a full likelihood algorithm, while the BMVS calculates the marginal likelihood, which requires less computational cost. Therefore, if the BMVS algorithm is able to be implemented in a more advanced platform such as the GPU like the GUESS does, it will also gain considerable improvement in the speed.}

The results of BMVS are quite robust when the number of response variables increases from $q = 5$ to $q = 30$ in simulation setting 2 or the number of influential predictors doubles from ten to twenty in simulation setting 4. At a minimum, this simulation study demonstrates the potential of the BMVS approach in modeling a 30-dimensional multivariate response vector. This provides response dimensionality large enough for many real life applications. Setting 3 represents a more difficult case because its influential signal is very weak and the correlations between predictors add further confounding factors. As a result, it is to be expected that the results of Setting 3 are a little worse than those of the other settings. We demonstrate that an initial feature screening, distance correlation based sure independence screening before applying BMVS (i.e., DC-SIS+BMVS), is able to improve the results \cite{li2012feature}. \tr{Fan and Lv \cite{fan2008sure} proposed the sure independence screening (SIS) procedure and proved that the Pearson correlation ranking procedure possesses a sure screening property. As an extension, Li et al. \cite{li2012feature} proposed the distance correlation-based sure independence screening procedure, which can be applied to pre-select important predictors based on their association strength with the multivariate response.}

The performance of the GUESS is comparable to that of the BMVS approach for the majority cases, and even performs better than BMVS in simulation setting 3. This maybe caused by the conservativeness of AICc in terms of selecting an optimal model size. As a result, the model selected by BMVS tends to be smaller than the true model in simulation setting 3, causing its power to be slightly lower than the other settings, but it does not imply that the posterior probabilities of the truly influential predictors generated by BMVS are also small. When a good sample size is given, BMVS significantly outperforms GUESS and accurately selects the exact true model 100\% times in simulation setting 4.

Scott and Berger \cite{scott2010bayes} discussed the multiplicity correction issues in Bayesian variable selection and claimed that no fixed choice of $\phi = P(Z_i = 1)$ that is independent of the sample size $n$ can adjust for multiplicity. We recommend choosing $\phi$ such that $P(\sum_{i = 1}^p Z_i > K) = 0.1$, where $K = \text{log}n$, and the $\phi$ value can be easily derived as a Binomial distribution parameter. After varying the noise-to-signal ratios in the simulation studies, we find that the recommended prior probability is robust to different settings.

\begin{table}[H]
\centering
\caption{\label{tab1.1}Simulation results for Setting 1.}
\begin{tabular}{l c c c c c c}\\
\hline
 & $mpp_1$ & $mpp_0$ & $\mathbb{P}(\hat{t} = t)$ & $\mathbb{P}(\hat{t} \supset t)$ & FDR  \\ \hline
 & \multicolumn{6}{c}{$n = 200, \quad p = 500, \quad q = 5, \quad |t| = 10$} \\\\
BMVS & 0.999 & $5.234 \times 10^{-14}$ & 0.990 & 0.990 & 0.000  \\
CCA & & & 0.390 & 0.960 & 0.085 & \\
M-LASSO & & & 0.000 & 1.000 & 0.949 & \\
M-EN & & & 0.000 & 1.000 & 0.948 & \\
R2GUESS & & & 0.990 & 1.000 & 0.001 & \\ \hline
 & \multicolumn{6}{c}{$n = 200, \quad p = 1000, \quad q = 5, \quad |t| = 10$} \\\\
BMVS & 1.000 & $1.215 \times 10^{-15}$ & 1.000 & 1.000 & 0.000  \\
CCA & & & 0.350 & 0.960 & 0.089 & \\
M-LASSO & & & 0.000 & 1.000 & 0.950 & \\
M-EN & & & 0.000 & 1.000 & 0.948 & \\
R2GUESS & & & 1.000 & 1.000 & \tr{0.000} & \\
\hline
\end{tabular}
\end{table}

\begin{table}[H]
\centering
\caption{\label{tab1.2} Simulation results for Setting 2.}
\begin{tabular}{l c c c c c c}
\hline
 & $mpp_1$ & $mpp_0$ & $\mathbb{P}(\hat{t} = t)$ & $\mathbb{P}(\hat{t} \supset t)$ & FDR  \\ \hline
 & \multicolumn{6}{c}{$n = 200, \quad p = 500, \quad q = 30, \quad |t| = 10$} \\\\
BMVS & 1.000 & $3.009 \times 10^{-105}$ & 1.000 & 1.000 & 0.000  \\
CCA & & & 0.280 & 1.000 & 0.088 & \\
M-LASSO & & & 0.000 & 1.000 & 0.945 & \\
M-EN & & & 0.000 & 1.000 & 0.941 & \\
R2GUESS & & & 1.000 & 1.000 & 0.000 & \\ \hline
 & \multicolumn{6}{c}{$n = 200, \quad p = 1000, \quad q = 30, \quad |t| = 10$} \\\\
BMVS & 1.000 & $2.015 \times 10^{-114}$ & 1.000 & 1.000 & 0.000  \\
CCA & & & 0.340 & 1.000 & 0.098 & \\
M-LASSO & & & 0.000 & 1.000 & 0.946 & \\
M-EN & & & 0.000 & 1.000 & 0.941 & \\ 
R2GUESS & & & 1.000 & 1.000 & 0.000 & \\
\hline
\end{tabular}
\end{table}

\begin{table}[H]
\centering
\caption{\label{tab1.3} Simulation results for Setting 3.}
\begin{tabular}{l c c c c c c}
\hline
 & $mpp_1$ & $mpp_0$ & $\mathbb{P}(\hat{t} = t)$ & $\mathbb{P}(\hat{t} \supset t)$ & FDR  \\ \hline
 & \multicolumn{6}{c}{$n =200, \quad p = 500, \quad q = 5, \quad |t| = 10$} \\\\
BMVS & \tr{0.995} & 0.001 & 0.750 & \tr{0.950} & \tr{0.001}  \\
DC-SIS+BMVS & & & 0.880 & 0.950 & 0.006 & \\
CCA & & & 0.300 & 0.990 & 0.116 & \\
M-LASSO & & & 0.000 & 1.000 & 0.949 & \\
M-EN & & & 0.000 & 1.000 & 0.948 & \\
R2GUESS & & & 0.990 & 1.000 & 0.001 & \\ \hline
 & \multicolumn{6}{c}{$n =200, \quad p = 1000, \quad q = 5, \quad |t| = 10$} \\\\
BMVS & \tr{0.996} & \tr{$8.081 \times 10^{-5}$} & \tr{0.930} & \tr{0.960} & \tr{0.003}  \\
DC-SIS+BMVS & & & \textbf{0.950} & \textbf{0.970} & 0.002 & \\
CCA & & & 0.270 & 1.000 & 0.134 & \\
M-LASSO & & & 0.000 & 1.000 & 0.950 & \\
M-EN & & & 0.000 & 1.000 & 0.948 & \\ 
R2GUESS & & & 0.940 & 0.950 & 0.002 & \\
\hline
\end{tabular}
\end{table}

\begin{table}[H]
\centering
\caption{\label{tab1.4} Simulation results for Setting 4.}
\begin{tabular}{l c c c c c c}
\hline
 & $mpp_1$ & $mpp_0$ & $\mathbb{P}(\hat{t} = t)$ & $\mathbb{P}(\hat{t} \supset t)$ & FDR  \\ \hline
 & \multicolumn{6}{c}{$n = 100, \quad p = 500, \quad q = 5, \quad |t| = 20$} \\\\
BMVS & \tr{0.8925} & \tr{$6.20 \times 10^{-14}$} & 0.000 & \tr{0.080} & \tr{0.028}  \\
CCA & & & 0.000 & 0.000 & 0.084 & \\
M-LASSO & & & 0.000 & 1.000 & 0.793 & \\
M-EN & & & 0.000 & 0.760 & 0.794 & \\
R2GUESS & & & 0.000 & 0.000 & 0.000 & \\ \hline
 & \multicolumn{6}{c}{$n = 500, \quad p = 500, \quad q = 5, \quad |t| = 20$} \\\\
BMVS & 1 & \tr{$3.74 \times 10^{-14}$} & \textbf{1.000} & \textbf{1.000} & 0.000  \\
CCA & & & 0.000 & 0.000 & 0.084 & \\
M-LASSO & & & 0.000 & 1.000 & 0.960 & \\
M-EN & & & 0.000 & 1.000 & 0.959 & \\ 
R2GUESS & & & 0.000 & 0.000 & 0.000 & \\
\hline
\end{tabular}
\end{table}

\section{Real data analyses}
In this section we apply our BMVS method to two real data examples in genome-wide association studies, each containing multiple response variables. 

\subsection{Rice shape data}
In this example, we analyze a dataset related to the shape of rice (\emph{Oryza sativa}). There are 3,254 SNPs genotyped for each of the 179 rice accessions, with missing genotypes imputed by Iwata et al. \cite{iwata2015genomic}. The SNPs do not contain heterozygous genotypes due to the inbreeding nature of \emph{Oryza sativa} \cite{zhao2011genome}. Thus, the SNPs here only have genotype $AA$ or $aa$. As a standard routine, SNPs with minor allele frequencies of less than 5\% are excluded in our analyses. Though using the same data, the previous work of Iwata et al. \cite{iwata2015genomic} mainly focused on shape prediction, which is different from our emphasis for identifying the most influential genetic variants.

We perform principal component analysis on the shape descriptor, EFD coefficients, reported by Iwata et al. \cite{iwata2015genomic}. The first six PCs are retained as the response of our model ($q=6$), which explain about 99\% of the total variation in the shape descriptor coefficients (the first six PCs explain 94.66\%, 1.85\%, 1.00\%, 0.57\%, 0.52\%, and 0.26\% of the total variation respectively). We then perform a five-fold cross-validation on the entire observed data and calculate the following numerical assessments: 1) average size of selected models ($|\hat{t}|$); 2) the Frobenius matrix norm of the difference between predicted phenotypes and observed phenotypes ($||\boldsymbol{Y} - \boldsymbol{\hat{Y}}||_F$); and 3) the multivariate AICc of selected models. The BMVS method is again compared with GUESS (\emph{R2GUESS}), M-LASSO (\emph{glmnet}), and M-EN (\emph{glmnet}).

Table \ref{tab1.5} shows the results of the five-fold cross-validation obtained using these four methods. BMVS method produces the lowest values of AICc ($724.6854$) among all of the four methods. The M-LASSO approach selects an average model size of 139, which is around 12 times larger than that of the BMVS, dramatically inflating both its prediction error and its AICc. Although M-EN yields the lowest prediction error, its AICs and model size are both large. The GUESS algorithm selects a slightly smaller model size than BMVS, but at the cost of increasing both AICc and prediction error. In addition to comparing BMVS with standard multivariate variable selection approaches, we also compare it with its univariate counterpart to assess whether a multivariate approach performs better than a univariate approach when handling multiple responses. Here ``Univariate counterpart'' refers to the process of individually fitting each component of the multivariate response using univariate approaches and repeating the process six times. To make comparison fair and also minimize all other possible confounding factors, we used the Bayesian variable selection method proposed by Narisetty and He \cite{narisetty2014bayesian} from which the BMVS is extended. This ``univariate counterpart'' method yields a predictor error $||\boldsymbol{Y} - \boldsymbol{\hat{Y}}||_F$ of $7.0076$, which is about three times larger than that yielded by BMVS. 

\begin{table}
\caption{\label{tab1.5} Results of real data analyses on rice shape data using five-fold cross-validation}
\centering
\begin{tabular}{l c c c}
\hline
& $|\hat{t}|$ & $||\boldsymbol{Y} - \boldsymbol{\hat{Y}}||_F$ & AICc \\ \hline
BMVS & 11.80 & 2.8626 & 724.6854 \\
R2GUESS & 8.40 & 6.0624 & 883.5024 \\
M-LASSO & 139.60 & 4.4521 & 9915.9830 \\
M-EN & 101.60 & 1.7921 & 3716.6670 \\
\hline
\end{tabular}
\end{table}

\begin{table}
\caption{\label{tab1.6} Results of real data analyses on flowering time data using five-fold cross-validation}
\centering
\begin{tabular}{l c c c}
\hline
& $|\hat{t}|$ & $||\boldsymbol{Y} - \boldsymbol{\hat{Y}}||_F$ & AICc \\ \hline
$\hat{t}_{\text{BMVS}}$ & 9 & 16.6450 & -179.2434 \\
$\hat{t}_{\text{1}}$ & 70 & 64.7921 & 227.8176 \\
$\hat{t}_{\text{2}}$ & 215 & 997.0443 & 5294.179 \\
\hline
\end{tabular}
\end{table}

\subsection{Flowering time data}
In this example, we consider the genome-wide association studies for three flowering time related response of 272 rice \emph{Oryza sativa} accessions: flowering time (FT) at Aberdeen, FT ratio of Arkansas/Aberdeen, and FT ratio of Faridpur/Aberdeen. These three phenotypes of the same rice accession are correlated, sharing genetic basis in some way. The predictor data includes 36,901 SNPs that were genotyped by Zhao et al. \cite{zhao2011genome} and imputed by Iwata et al. \cite{iwata2015genomic}. For this ultrahigh-dimensional setting, we perform the DC-SIS+BMVS process. Specifically, we keep 200 candidate SNPs from the initial feature screening and then perform BMVS on this subset to further detect the truly influential SNPs.

Zhao et al. \cite{zhao2011genome} conducted genome-wide association studies for the same dataset but used linear mixed model to fit each individual response separately. They reported significant SNPs ($p-\text{value} < 10^{-4}$) for each individual phenotype. Here, we compare three sets of results as follows: 1) the SNPs selected by the BMVS method, denoted by $\hat{t}_{\text{BMVS}}$; 2) the SNPs that are simultaneously significant for all three phenotypes as reported by Zhao et al. \cite{zhao2011genome}, denoted by $\hat{t}_{\text{1}}$ (the intersection set); 3) the SNPs that are significant for at least one phenotype as reported by Zhao et al. \cite{zhao2011genome}, denoted by $\hat{t}_{\text{2}}$ (the union set).

\begin{figure}[t]
\centering
  \centerline{\includegraphics[width=0.8\textwidth]{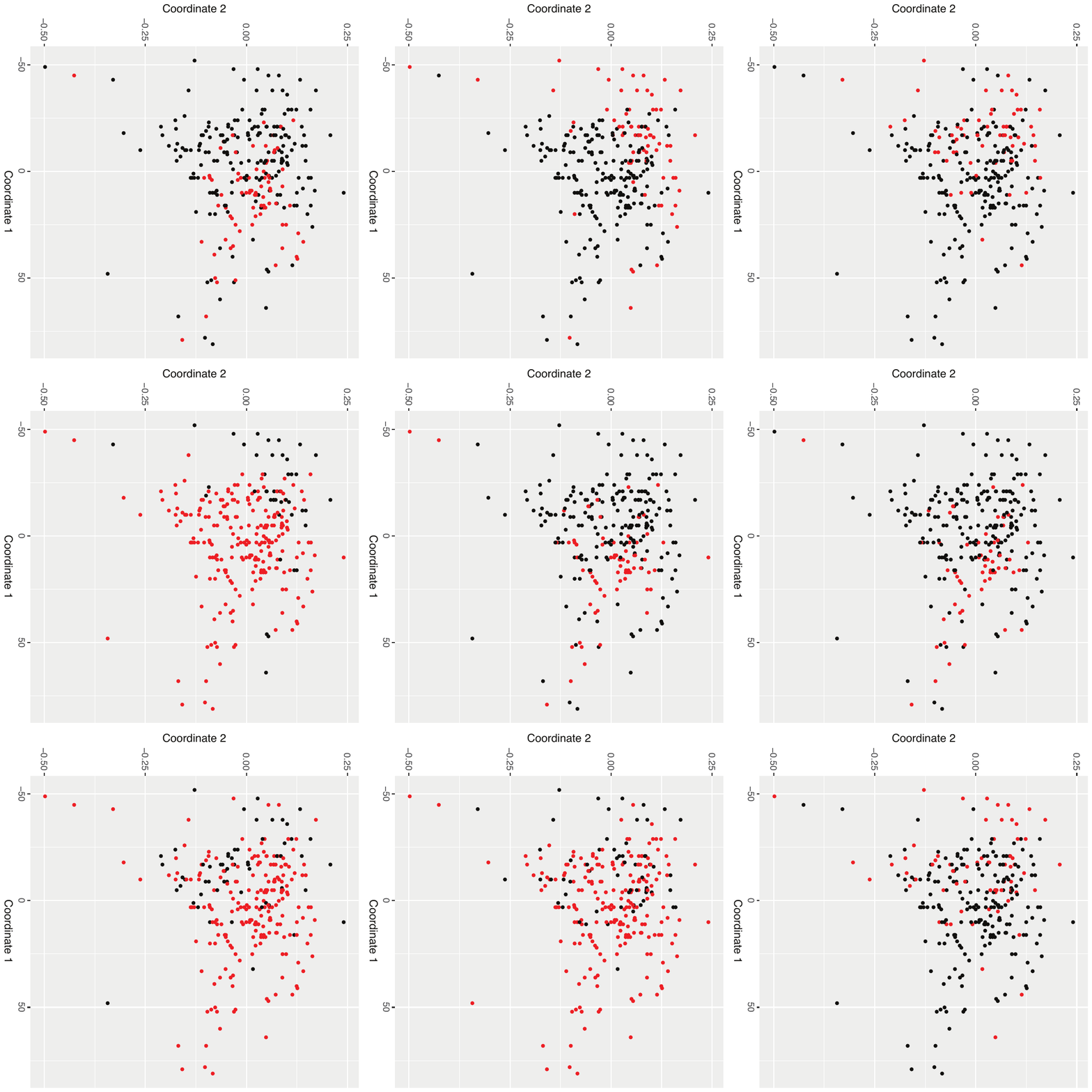}}
  \caption{The multidimensional scaling plot demonstrating variation of the multivariate response vector under different genotype groups in a two-dimensional space for each of the nine influential SNPs selected by BMVS approach,  with black dots representing genotype aa and red dots representing genotype AA.}
  \label{fig1.2}
\end{figure}

The BMVS method selects 9 SNPs, the intersection set $\hat{t}_{\text{1}}$ contains 70 SNPs, and the union set $\hat{t}_{\text{2}}$ contains 215 SNPs. To perform a fair comparison, we apply a multivariate linear regression model to perform five-fold cross-validation for these three sets of results (see Table \ref{tab1.6}). BMVS outperforms the other methods by using the smallest model size to achieve the best results. Specifically, it achieves both the smallest prediction error (16 versus 64 or 997) and the smallest AICc (-179 versus 227 or 5294). The negative sign of the AICc obtained by BMVS may cause some confusion, but the AICc rules indicate smaller results are always more desirable than larger ones. Compared to $\hat{t}_{\text{BMVS}}$, $\hat{t}_{\text{2}}$ is astonishingly large in both selected model size and the predictor error. In addition, we further compare the significance of the three sets by directly fitting a multivariate regression model that use all observations (without doing cross validation). Eight out of nine selected SNPs are significant for $\hat{t}_{\text{BMVS}}$, while only 10 out of 70 selected SNPs are significant for $\hat{t}_{\text{1}}$ and only 52 out of 215  significant for $\hat{t}_{\text{2}}$. All these analyses are made at the 0.05 level without adjustment for multiple comparisons.

To visualize the associations between the 3-dimensional response vector and each of the nine influential SNPs selected by BMVS (i.e., set $\hat{t}_{\text{BMVS}}$), we use the multidimensional scaling (MDS) approach to demonstrate the Euclidean distances of 272 observations in a two-dimensional space (see Figure \ref{fig1.2}) \cite{kruskal1978multidimensional}. The MDS approach calculates the pairwise Euclidean distance matrix of all the observations and extract eigenvectors associated with the top two eigenvalues. This ensures that the pairwise Euclidean distances can be approximated and visualized in a lower dimensional space (see Figure \ref{fig1.2}). In each subplot of Figure \ref{fig1.2}, the color coding represents different genotypes for each of the nine influential SNPs, with black dots for genotype $aa$ and red dots for genotype $AA$. Apparently, the variation of the response vector shows quite different patterns of distribution under two different genotypes of each of the nine influential SNPs, which may indicate that the associations are not trivial.

\section{Discussion}
One appealing characteristic of the BMVS method is that it skips 
the mundane process of tuning parameter selections, as required by penalized regression approaches. All simulation and real data examples use the same value for 
hyperparameters with no additional tunings required. Another appealing property is that the theoretical selection consistency (see supplementary material) holds for ultrahigh dimensional cases, including situations where the number of predictors is exponentially larger than the sample size (i.e., $p = e^{o[n]}$). 

In general, Bayesian methods are more computationally expensive than the frequentist approaches due to their large number of hyperparameters and their long iteratively updating process. Bhattacharya et al. \cite{bhattacharya2016} and Narisetty et al. \cite{narisetty2018} proposed fast sampling methods for high-dimensional multivariate Gaussian distributions, which may be exploited to improve the speed for BMVS. For the ultrahigh dimensional data such as genome-wide association studies that we analyzed in this article, we suggest using a fast initial feature screening method before applying the BMVS method. As one can see from the results of our simulation 3 and flowering time data analyses, the combination of DC-SIS and BMVS improves the accuracy.


Many studies \ta{modeled a multivariate response by separately fitting each component of the response vector using multiple univariate models \cite{zhang2009penalized, chitwood2014resolving, narisetty2014bayesian, fu2017quantitative}. Compared to the BMVS approach, which handles the multivariate response as one unit, the separate univariate regression approach has two notable disadvantages:} 1) They fail to take advantage of the correlations between multiple response variables \cite{breiman1997predicting}; \ta{but correlated variables indeed share information in a way that is important for discovering additional signals} \cite{o2012multiphen}.
2) When a series of models are fitted, deriving a generalized interpretation for multiple response variables is difficult \cite{chitwood2014resolving, fu2017quantitative}. 3) It is more appropriate to group multiple correlated response variables together for some practical applications, and hence separating them could produce incomplete or misleading results. Breiman and Friedman \cite{breiman1997predicting} provided theoretical proofs demonstrating that prediction accuracy can be improved when modeling multiple correlated response variables by using one multivariate regression model compared to using multiple separate univariate regression models. This claim is empirically confirmed in two of our real data analyses.

For the future work, we may consider an improved approach to determine the optimal model size, i.e., the threshold to determine how many number of predictors to keep. Even though this article primarily focuses on variable selection, the BMVS can also be considered for prediction purpose. If prediction is the main interest, one can use other criteria such as cross-validation or bootstrap prediction error in place of the multivariate AICc to improve the prediction accuracy.

\section{Appendix}\label{section}

\tO{
\ta{The notations are defined as follows}:

\begin{itemize}
\item Let $k$ be a $p\times 1$ binary vector that represents an arbitrary model. The $i$th entry, $k_i$, of $k$ indicates whether  the $i$th predictor is active (indicated by
$k_i = 1$) or inactive ($k_i = 0$) in the model. We denote the true model by $t$. We use $|k| = k_1 + k_2 + \cdots + k_p$ to indicate the size of model $k$.
Given two models $k$ and $j$, the operations $k \land j$ and $k \lor j$ denote the entry-wise minimum and maximum, respectively.
Also, define $k^{c} := \bm{1} - k$ to be the complement model to $k$, where $\bm{1}$ represents the $p$-dimensional vector of 1's.
We also use the notation $k \supset j$ to mean that model $k$ includes all predictors in model $j$. Otherwise, $k \not\supset j$ is used in the case where $j$ contains some predictor not in $k$.
\item Let $\bm{X}_k$ refer to the matrix of predictors retained by the binary vector $k$. Then $\covkXXk{k}/nq$ is the Gram matrix for $\Xk{k}$. Similarly, let $\bm{\beta}_k$ be the matrix of coefficients associated with the model $k$.
\item For any $\psi > 0$, define \[m_n(\psi) := \min\left(p, \frac{nq}{(2+\psi)\log p}\right).\]
\item Let $\zeta(A)$ be the function on a square matrix that returns the minimum non-zero eigenvalue of $A$.
Then define \[\lambda_{min}(\psi) := \inf_{|k|\leq m_n(\psi)} \zeta\left(\frac{\covkXXk{k}}{nq}\right). \]
\item We also define $\lambda_{max}$ to be the maximum eigenvalue of the full Gram matrix $\covkXXk{}/nq$ (note here that we are using the full Gram matrix). Further note that for any desirable model, the minimum eigenvalue of $\covkXXk{k}/nq $ will always be non-zero; a model with an associated Gram matrix that is singular will suffer from issues of collinearity.
\item For square matrices $A$ and $B$ of the same dimension, we use $A \geq B$ to denote that $A-B$ is a positive semi-definite matrix.
\item Finally, a comment on further notation is in order. For real valued sequences $a_n$ and $b_n$, $a_n \sim b_n$ means that $\frac{a_n}{b_n} \to c >0$, $a_n \preceq b_n$ means $a_n = \mathcal{O}(b_n)$, and $a_n \prec b_n$ means $a_n = o(b_n)$.
\end{itemize}

\subsection*{Conditions}
Under the model given by Equation (2) of the main text, we provide a collection of conditions that we will assume throughout the remainder of this paper. As many meaningful applications of the BMVS method are in the context of $p \gg n$ (i.e. a high or ultrahigh dimensional feature space), we will tacitly assume $p \rightarrow \infty$  as $n\rightarrow \infty$. This leads us to our first condition:

\subsubsection*{Condition 1} $p = e^{n \xi}$ for some $\xi \to 0$ as $n \to \infty$. In other words, $\log p = o(n)$. This restricts the number of predictors to be no greater than exponential in $n$. \textcolor{black}{The dimension $q$ of $\bm Y$, instead, is a fixed positive constant}. The remaining conditions are as follows.

\subsubsection*{Condition 2} \textcolor{black}{$n\tau^2_0 = o(1)$, $n\tau^2_1 \sim \max(n, p^{2 + 3\delta})$, for some $\delta > 0$}. We also assume $\phi \sim p^{-1}$. This condition establishes shrinkage and diffusion rates for the spike ($\tau_0^2$) and slab ($\tau_1^2$) priors.

\subsubsection*{Condition 3} (A condition on the true model). $\bm{Y} | \bm{X} \sim N_q(\bm{X}_t\bm{\beta}_t + \bm{X}_{t^c}\bm{\beta}_{t^c},~ \bm{\Sigma}_Y)$, with $|t|$, the size of the true model, being fixed. Note that $\bm{\beta}_{t^c}$ can have non-zero components, but yet must satisfy $\vertiii{\bm{X}_{t^c}\bm{\beta}_{t^c}} = \mathcal{O}(1),$ where we take $\vertiii{\cdot}$ to be the Frobenius matrix norm. It should be noted that the matrix norm on $M(\mathbb{R})$ (an arbitrary ring of matrices over $\mathbb{R}$) induced by the $\bm{\ell}_2$ norm on $\mathbb{R}^2$ can also be used here; however, the induced norm will always yield a stricter condition, since the matrix norm induced by the  $\bm{\ell}_2$ norm is always bounded by the Frobenius norm (this is a classical result). For a formal discussion on this matter, see Theorem 5.6.34 in \cite{horn2012matrix} (while noting that the Frobenius norm is unitarily invariant).


Let $K\in \mathbb{R}$ be fixed. Define the following:
\[\Delta_n(K):= \inf_{\{k:|k| < K|t|, k \not\supset t\} }\left(\vertiii{(\bm{I} - \bm{P}_k)\bm{X}_{t}\bm{\beta}_t}\right)^2,\]
where $\bm{P}_k$ represents the standard ordinary least squares projection matrix onto the column space of $\bm{X}_k$, with $k$ varying. This leads to another desirable condition:
\subsubsection{Condition 4} (Identifiability). There is $K > 1 + 8/\delta$ such that $\Delta_n(K) > \gamma_n:= 5|t|(1 + \delta)\log\left(\max(\sqrt{nq}, p)\right)$, for the $\delta > 0$ used in condition 2.

\subsubsection*{Condition 5} (Design regularity). For some $\psi < \delta$, $\kappa < (K -1)\delta/2$,
\[\lambda_{max} \prec \min(1/(nq\sigtau{0}^2), nq \sigtau{1}^2),\] \[ \text{and} ~~\lambda_{min}(\psi) \succeq \max\left(\frac{\max\left(nq, p^{2( 1+ \delta)}\right)}{nq\sigtau{1}^2}, p^{-\kappa}\right).\]


We now turn our attention to the main theoretical results of our paper. 

\subsection*{Main Lemma and Theorem}
To simplify notation, we will suppress $\psi$ from $\lambda_{min}(\psi) $ and $m_n(\psi)$. Given a model $k$ along with the true model, define the following probability ratio (PR):

\[\text{PR}(k,t) := \frac{\mathbb{P}(\bm{Z} = k | \bm{Y}, \sigma_\beta^2, \bm{\Sigma}_Y)}{\mathbb{P}(\bm{Z} = t | \bm{Y}, \sigma_\beta^2, \bm{\Sigma}_Y)}\]

This is the posterior ratio of a model $k$ with respect to the true model $t$. We now need to establish a collection of notation before proceeding forward. Define the following:
\[D_k = \text{Diag}(k(\sigma_\beta\tau_1)^{-2} + (\bm{1} - k)(\sigma_\beta\tau_0)^{-2}),\]
\[\mathcal{W}_k = \covkXXk{} + (\bm{I}_q \otimes D_k),\]
\[ Q_k = |\mathcal{W}_k|^{-1/2}|\bm{I}_q \otimes D_k|^{1/2}, \quad s_n = \frac{\phi}{1-\phi},\]
\[r_k = \text{rank}(\bm{X}_k), \quad r_k^{*} = \min(r_k, m_n),\]
and
\[\quad \widetilde{R}_k = \vec{\bm{Y}}' \bm{\Sigma}_Y^{0} \vec{\bm{Y}}.\]
where $\bm{\Sigma}_Y^{0} = \left((\bm{\Sigma}_Y^{-1} \otimes \bm{I}_n) - (\bm{\Sigma}_Y^{-1} \otimes \bm{X})\mathcal{W}_k^{-1}(\bm{\Sigma}_Y^{-1} \otimes \bm{X}')\right)$.
Here $\widetilde{R}_k$ is the residual sum of squares for a shrinkage estimator of $\bm{\beta}$ under the premise of generalized least squares estimation. With this generalized definition of $\widetilde{R}_k$, we now have an important lemma.
Taking the assumptions of Conditions 2 and 5, we can state the following for any model $k \neq t$:
\begin{lemma}
\textcolor{black}{For any fixed $\sigma_\beta^2$, $\Sigma_Y$, we have}
\begin{eqnarray*}\label{LEMMA}
\text{PR}(k,t) &=& \frac{Q_k}{Q_t}s_n^{|k| - |t|}\text{exp}\left\{-\frac{1}{2}(\widetilde{R}_k - \widetilde{R}_t)\right\}\\\\
&\leq & w\lambda_{min}^{-|k^c \land t|/2}(nq\sigtau{1}^2\lambda_{min}(1-h))^{(r_k^* - r_{t})/2}\\\\
&\times & s_n^{|k| - |t|} \text{exp}\left\{-\frac{1}{2}(\widetilde{R}_k - \widetilde{R}_t)\right\},
\end{eqnarray*}
where $w > 0$ is a constant and $h  = o(1)$.
\end{lemma}

The proof of Lemma 1 is addressed in the Supplement file. Note that the term $\text{LR} := \text{exp}\{-\frac{1}{2}(\widetilde{R}_k - \widetilde{R}_t)\}$ corresponds to the standard likelihood ratio of the model $k$ versus the true model. First consider a model that is missing one or more ``true'' predictor. Then the term $\widetilde{R}_k - \widetilde{R}_t$ will go to infinity at the same rate as $n$, since this is approximately the residual sum of squares between $k$ and the true model. It then easily follows that $\text{PR}(k,t) \to 0$ (since the exponential in $\text{PR}(k,t)$ is the clear dominant term based on the aforementioned conditions).
Now let us consider the case where $k \supset t$ and $|k| > |t|$ (so $k$ contains all ``true'' predictors, as well as one or more noise predictors). In this case, $ \widetilde{R}_k - \widetilde{R}_t$ is probabilistically bounded. Furthermore, since the non-exponential term of $\text{PR}(k,t)$ goes to zero in this case, $\text{PR}(k,t)$ itself goes to zero. Note that when $r_k > r_t$, this non-exponential term of $\text{PR}(k,t)$ becomes reciprocally smaller as $\tau_1^2$ increases. Hence, the posterior ratio for larger models goes to zero more quickly for larger values of $\tau_1^2$. 

\begin{thm}[Strong Sure Selection]\label{theorem1}
\textcolor{black}{Assume Conditions 1 through 5. Under the structure of model (2), we have \[\mathbb{P}(\bm{Z} = t | \bm{Y}, \bm{X}) \xrightarrow{P} 1,\] as $n \to \infty$ under the true data distribution, where $\bm{Z}$ is the 0,1 vector representing the model selected by the BMVS method and $t$ is the true model.}
\end{thm}

Under relatively weak conditions, even misspecified $\sigma_\beta^2$ or $\bm{\Sigma}_Y$ still yield equivalent results. It is suggested that readers familiarize themselves with the \textbf{vec} operator and the Kronecker product, as well as the concepts of generalized multivariate least squares regression. These topics are used extensively in our proofs.}




\tO{
\section*{Proof of Lemma 1}\label{lemmaProof} We will herein prove Lemma 1. 
\subsection{Proof} \textcolor{black}{As given in Section 2 of the main text, $\bm{X}$ is independent with $\bm{\beta}, \bm{\Sigma}_Y$ and $\sigma_\beta^2$, the joint posterior of $\bm{\beta}, \bm{Z}, \sigma_\beta^2$, and $\bm{\Sigma}_Y$ given the data under our model at (2) is given by}
\small
\begin{align*}
&f(\bm{\beta}, \sigma_\beta^2, \bm{\Sigma}_Y,\bm{Z}| \bm{Y},\bm{X}) \propto f(\bm{Y}|\bm{X}, \bm{\beta}, \sigma_\beta^2, \bm{\Sigma}_Y,\bm{Z}) \times \prod_{i =1}^p f(\bm{\beta}_i)\prod_{i = 1}^p f(\bm{Z}_i|\bm{X})\\
&\quad\quad\quad\quad\quad\quad\quad\quad\quad \times f(\sigma_\beta^2) \times f(\bm{\Sigma_Y})
\end{align*}
\normalsize
First let us examine just $f(\bm{Y}|\bm{X}, \bm{\beta}, \sigma_\beta^2, \bm{\Sigma}_Y,\bm{Z})$. Based on what we established in Section 2, as well as through further algebraic arrangement, we now can state the following

\begin{align*}
&f(\bm{Y}|\bm{X}, \bm{\beta}, \sigma_\beta^2, \bm{\Sigma}_Y,\bm{Z})\\\\
&\propto \left(\frac{1}{\sqrt{|\bm{\Sigma}_Y|}}\right)^n\text{exp}\left\{-\frac{1}{2}\sum_{m = 1}^n (\bm{Y}_m - \bm{X}_m\bm{\beta})\bm{\Sigma}_Y^{-1} (\bm{Y}_m - \bm{X}_m\bm{\beta})^{\prime} \right\}\\\\
&\propto \left(\frac{1}{\sqrt{|\bm{\Sigma}_Y|}}\right)^n \text{exp}\bigg\{-\frac{1}{2}\left[\vec{\bm{Y}}'(\bm{\Sigma}_Y^{-1} \otimes \bm{I}_n)\vec{\bm{Y}} \right.\\
& \left. \quad\quad\quad\quad\quad\quad\quad\quad\quad - ~2\vec{\bm{Y}}'(\bm{\Sigma}_Y^{-1} \otimes \bm{X})\vec{\bm{\beta}}\right.\\
& \left. \quad\quad\quad\quad\quad\quad\quad\quad\quad + ~\vec{\bm{\beta}}'\covkXXk{}\vec{\bm{\beta}}\right]\bigg\}.
\end{align*}
\normalsize

It is a simple exercise to verify that we now have \textit{in total} the following for the joint posterior given by \textcolor{black}{$f(\bm{\beta},\bm{Z}, \sigma_\beta^2, \bm{\Sigma}_Y| \bm{Y},\bm{X}):$}
\small
\begin{align*}
&f(\bm{\beta}, \sigma_\beta^2, \bm{\Sigma}_Y,\bm{Z}| \bm{Y},\bm{X})\\ &\propto \left(\frac{1}{\sqrt{|\bm{\Sigma}_Y|}}\right)^n
\text{exp}\bigg\{-\frac{1}{2}\left[\vec{\bm{Y}}'(\bm{\Sigma}_Y^{-1} \otimes \bm{I}_n)\vec{\bm{Y}} \right.\\
& \left. \quad\quad\quad\quad\quad\quad\quad\quad\quad - ~2\vec{\bm{Y}}'(\bm{\Sigma}_Y^{-1} \otimes \bm{X})\vec{\bm{\beta}}\right.\\
& \left. \quad\quad\quad\quad\quad\quad\quad\quad\quad + ~\vec{\bm{\beta}}'\covkXXk{})\vec{\bm{\beta}}\right.\\
& \left. \quad\quad\quad\quad\quad\quad\quad\quad\quad + ~\vec{\bm{\beta}}'(\bm{I}_q\otimes D_k)\vec{\bm{\beta}}-{2\alpha_2}/{\sigma_{\beta}^2}\right]\bigg\}\\
&\quad\quad\quad\quad\quad\quad\quad\quad\quad \times \quad\sigma_{\beta}^{-2(nq/2 + \alpha_1 + 1)}|\bm{\Sigma_Y}|^{-(\nu + q + 1)/2}\\
&\quad\quad\quad\quad\quad\quad\quad\quad\quad \times \quad\text{exp}\left\{\text{tr} (\Lambda\bm{\Sigma_Y}^{-1})\right\}|\bm{I}_q \otimes D_k|^{1/2}s_n^{|k|},
\end{align*}
\normalsize
where $D_k$ and $s_n$ are as defined before. Here $\alpha_1$ and $\alpha_2$ are the shape and scale parameters, respectively, of the inverse gamma prior of $\sigma_\beta^2$. Additionally, $\nu$ and $\Lambda$ are the parameters of the prior distribution (inverse Wishart) on $\bm{\Sigma}_Y$.

Define a matrix $\bm{\tilde{\beta}}$ as follows:
\[\vec{\tilde{\bm{\beta}}} = \left(\covkXXk{} + (\bm{I}_q \otimes D_k)\right)^{-1} (\bm{\Sigma}_Y^{-1} \otimes \bm{X}')\vec{\bm{Y}}.\]
This can be written more simply as
\[\vec{\bm{\tilde{\beta}}} = \mathcal{W}_k^{-1} (\bm{\Sigma}_Y^{-1} \otimes \bm{X}')\vec{\bm{Y}}.\]
By rearranging the terms in the above expression and dropping as constant the terms only involving known parameters, \textcolor{black}{we now can obtain the following}:
\begin{align*}
&f(\bm{\beta}, \bm{Z} = k | \bm{Y}, \bm{X}, \sigma_\beta^2, \bm{\Sigma_Y}) \\
&\propto \text{exp}\left\{-\frac{1}{2}\left(\vec{\bm{Y}}'(\bm{\Sigma}_Y^{-1} \otimes \bm{I}_n)\vec{\bm{Y}} \right.\right.\\
&\quad\quad\quad\quad\quad +\vec{\bm{\beta} - \bm{\widetilde{\beta}}}^{\prime}\mathcal{W}_k\vec{\bm{\beta} - \bm{\widetilde{\beta}}} \\
&\left.\left.\quad\quad\quad\quad\quad - ~\vec{\bm{\widetilde{\beta}}}^\prime\mathcal{W}_k \vec{\widetilde{\bm{\beta}}}\right)\right\}\times |\bm{I}_q \otimes D_k|^{1/2}s_n^{|k|}.
\end{align*}
Note that $\widetilde{\bm{\beta}}$ is a shrinkage estimator of the coefficient matrix $\bm{\beta}$. Given $\bm{Z} = k$, $(\bm{I}_q \otimes D_k)$ here acts as the precision matrix of $\bm{\beta}$ (the coefficient matrix). Shrinkage of each $\bm{\beta}_i$ will depend on $D_k$. When $k_i = 0$, then the components of $\widetilde{\bm{\beta}}_i$ shrink to zero. Furthermore, since $\tau_1^2$ is small, the shrinkage of coefficients in $\widetilde{\bm{\beta}}_i$ which correspond to $k_i = 1$ becomes negligible.

We now wish to use the joint posterior of $\bm{\beta}$ and $\bm{Z}$ to find the marginal posterior of just $\bm{Z}$. In order to do this, \textcolor{black}{we must compute the integral below}:
\begin{equation}\label{jointIntegral}
f(\bm{Z} = k | \bm{Y}, \bm{X}, \sigma_\beta^2, \bm{\Sigma_Y}) = \int_{\mathbb{R}^{p \times q}} f(\bm{\beta}, \bm{Z} = k | \bm{Y}, \bm{X}, \sigma_\beta^2, \bm{\Sigma_Y})d\bm{\beta}
\end{equation}

Using what we know about $f(\bm{\beta}, \bm{Z} = k | \bm{Y}, \bm{X}, \sigma_\beta^2, \bm{\Sigma_Y})$ we can state the following in relation to the integral described at (\ref{jointIntegral}):
\begin{align*}
&f(\bm{Z} = k | \bm{Y}, \bm{X}, \sigma_\beta^2, \bm{\Sigma_Y}) \\
&\propto  \text{exp}\{-\frac{1}{2}(\vec{\bm{Y}}'(\bm{\Sigma}_Y^{-1} \otimes \bm{I}_n)\vec{\bm{Y}} \\
& \quad\quad\quad\quad - \vec{\bm{\tilde{\beta}}}^\prime\mathcal{W}_k \vec{\bm{\tilde{\beta}}})\} \\
& \quad\quad\quad\quad \times |\bm{I}_q \otimes D_k|^{1/2}s_n^{|k|} \times \mathfrak{T}_k,
\end{align*}
where $\mathfrak{T}_k$ is defined by
\[\mathfrak{T}_k := \int_{\mathbb{R}^{p \times q}} \text{exp}\left\{-\frac{1}{2}\left(\vec{\bm{\beta} - \bm{\widetilde{\beta}}}^{\prime}\mathcal{W}_k\vec{\bm{\beta} - \bm{\widetilde{\beta}}} \right)\right\}d\bm{\beta}\]
Note now, however, that after the routine change of variables $u = \vec{\bm{\beta} - \bm{\widetilde{\beta}}}$,  $\mathfrak{T}_k$ is the standard multivariate Gaussian integral
\[\int_{\mathbb{R}^{p \times q}}\text{exp}\left\{-\frac{1}{2}\left(u'\mathcal{W}_ku\right)\right\}du.\]
(Of importance here, remember that $\mathcal{W}_k$ is a symmetric matrix). It is a standard result (see for example \cite{zinnjustin1993} and \cite{lancaster2014}) that
\[\int_{\mathbb{R}^{p \times q}}\text{exp}\left\{-\frac{1}{2}\left(u'\mathcal{W}_ku\right)\right\}du = \sqrt{\frac{(2\pi)^{pq}}{|{\mathcal{W}_k}|}}\]

Before proceeding, we define a generalized ``hat" matrix $\mathcal{H}_k$, as is often spoken of  in linear regression settings:
\[\mathcal{H}_k:= (\bm{\Sigma}_Y^{-1} \otimes \bm{I}_n)(\bm{I}_q \otimes X)\mathcal{W}_k^{-1}(\bm{I}_q \otimes X')(\bm{\Sigma}_Y^{-1} \otimes \bm{I}_n).\]
The matrix $\mathcal{H}_k$ is the hat matrix associated with the generalized least squares estimator of $\bm{\beta}$, yet with the inclusion of the precision matrix $(\bm{I}_q \otimes D_k)$ in the matrix $\mathcal{W}_k$.

We now can make the following statement:
\begin{align*}
&f(\bm{Z} = k | \bm{Y}, \bm{X}, \sigma_\beta^2, \bm{\Sigma}_Y)\\\\
&\propto Q_ks_n^{|k|}\:\text{exp}\{ -\frac{1}{2}(\vec{\bm{Y}}'(\bm{\Sigma}_Y^{-1} \otimes \bm{I}_n)\vec{\bm{Y}}\\
&\quad\quad\quad\quad\quad\quad\quad\quad - \vec{\bm{\tilde{\beta}}}^\prime\mathcal{W}_k \vec{\bm{\tilde{\beta}}})\rbrace\\\\
&= Q_ks_n^{|k|}\:\text{exp}\left\{ -\frac{1}{2}\bigg(\vec{\bm{Y}}^\prime\left((\bm{\Sigma}_Y^{-1} \otimes \bm{I}_n) - \mathcal{H}_k\right)\vec{\bm{Y}} \bigg)\right\rbrace\\\\
&=Q_ks_n^{|k|}\:\text{exp}\left\{ -\frac{1}{2}\widetilde{R}_k\right\rbrace,
\end{align*}
where $\widetilde{R}_k$ and $Q_k$ are still defined as previously. Thus it then follows that
\[\text{PR}(k,t) =\frac{Q_k}{Q_t}s_n^{|k| - |t|}\text{exp}\left\{-\frac{1}{2}(\widetilde{R}_k - \widetilde{R}_t)\right\}. \]
We now will obtain bounds on $Q_k$.

\subsection*{Lemma for finding bounds on $Q_k$}
In order to complete the proof of Lemma 1 we establish the following lemma. This lemma originates in \cite{narisetty2014bayesian}, however we now must adjust it to account for the covariance matrix $\bm{\Sigma}_Y$ and the definitions of $\widetilde{R}_k$ and $\mathcal{W}_k$.

Note that $\Sigma_Y^{-1}$ is a symmetric positive semi-definite matrix and hence is also unitarily diagonalizable. This in turn implies that there exists a well defined square root $\Sigma_Y^{-1/2}$ for $\Sigma_Y^{-1}$ (see, for example, Chapter 8 of \cite{bernstein2005matrix}).

\begin{lemma}\label{lemma111}
Let $A$ be an invertible matrix, and $B$ be any matrix with the appropriate dimension. Let $k$ and $j$ be any pair of models. We then have the following;
\begin{itemize}
\item[(i)] $|(A + B'B)^{-1}A| = |I  + BA^{-1}B'|^{-1}$
\item[(ii)] $\left(\bm{I}_{nq} + \sigtau{1}^2 \covXkkX{k} + \sigtau{0}^2\covXkkX{j}\right)^{-1}$ \\$~~~~~~~~~~~~~~~~~~~~~~~~~\geq \left(\bm{I}_{nq} + \sigtau{1}^2\covXkkX{k}\right)^{-1}(1-\xi_n),$ where $\xi_n = nq\sigtau{0}^2\lambda_{max}$.
\item[(iii)] $Q_k \leq w(nq\sigtau{1}^2\lambda_{min}(1-h))^{-(r_k^*-r_t)/2}\\ \times (\lambda_{min})^{-|t \land k^c|/2}Q_t$, \text{where} $w > 0$,\\
 $r_k = \text{rank}\left(\Xk{k}\right)$, $\min(r_k,m_n)$ and $h = {o}(nq)$.
\end{itemize}
\end{lemma}

By use of Lemma \ref{lemma111}, we now have
\begin{eqnarray*}\text{PR}(k,t) &=& \frac{Q_k}{Q_t}s_n^{|k|- |t|}\text{exp}\left\{-\frac{1}{2}(\widetilde{R}_k - \widetilde{R}_t)\right\}\\\\
&\leq & w\lambda_{min}^{-|k^c \land t|/2}(nq\sigtau{1}^2\\\\
&& \quad\quad \times \lambda_{min}(1-h))^{(r_k^* - r_{t})/2} s_n^{|k| - |t|}\\\\
&& \quad\quad \times \text{exp}\left\{-\frac{1}{2}(\widetilde{R}_k - \widetilde{R}_t)\right\}.
\end{eqnarray*}
This completes the proof of Lemma 1.\\
}

\section*{Proofs of Lemma \ref{lemma111}}

We prove each statement of Lemma \ref{lemma111} individually.
\begin{itemize}
\item[(i)] Remember that the determinant is multiplicative and that for any invertible matrix $C$, we know that $|C^{-1}| = |C|^{-1}$. We now have the following for invertible $A$ and $B$ being any matrix with the appropriate dimension:
\begin{eqnarray*}
|(A + B'B)^{-1}A| &=& |A + B'B|^{-1}|A|\\
&=& |A^{1/2}||A + B'B|^{-1}|A^{1/2}|\\
&=&|A^{-1/2}AA^{-1/2} + A^{-1/2}B'BA^{-1/2}|^{-1}\\
&=&|I + A^{-1/2}B'BA^{-1/2}|^{-1}\\
&=&|I + BA^{-1/2}A^{-1/2}B'|^{-1}\\
&=&|I + BA^{-1}B'|^{-1},
\end{eqnarray*}
where the second to last equality comes by Sylvester's determinant theorem.
This establishes the claimed result for (i).

\item[(ii)] We will use the Sherman-Morrison-Woodbury (SMW) matrix formula (See, for example, Fact 2.12.2 in \cite{bernstein2005matrix}). Let $A$, $B$, $C$, $D$ be matrices of conformable sizes such that $(C^{-1} + DA^{-1}B)$ and $(A + BCD)$ are both invertible. We then know that (this is again by SMW)
\[(A + BCD)^{-1} = A^{-1} - A^{-1}B(C^{-1} + DA^{-1}B)^{-1}DA^{-1}.\]
In reference to the above equation, let $A = (\bm{I}_{nq} + \sigtau{1}^2\covXkkX{k})$, $B = (\bm{\Sigma}_Y^{-1/2} \otimes X_j)$, $C = \bm{I}_{|j|q}$, and $D = (\bm{\Sigma}_Y^{-1/2} \otimes X_j')$. Now take any vector $\alpha$ of length $nq$. We then have
\begin{align*}
&\alpha'(A + BCD)^{-1}\alpha\\
&= \alpha'\left(\bm{I}_{nq} + \sigtau{1}^2\covXkkX{k} + \sigtau{0}\covXkkX{j}\right)^{-1}\alpha\\
&= \alpha'A^{-1}\alpha - \sigtau{0}^2H,
\end{align*}
where \[H = \vartheta'\left(\bm{I}_{|j|q} + \sigtau{0}^2\Xkp{j}A^{-1}\Xk{j}\right)^{-1}\vartheta,\]
\normalsize
with $\vartheta = \Xkp{j}A^{-1}\alpha$.
Since the matrix inside the quadratic form that is $H$ is a Gram matrix (and hence positive semi-definite by Theorem 7.2.10 of \cite{horn2012matrix}), we know $H \geq 0$ itself.
Thus we have the following:
\begin{eqnarray*}
0 &\leq&\sigtau{0}^2H\\
&\leq& \sigtau{0}^2\alpha'A^{-1}\covXkkX{j}A^{-1}\alpha\\
&\leq& nq\sigtau{0}^2\lambda_{max}\alpha'A^{-1}\alpha.
\end{eqnarray*}

\normalsize
Note that this last inequality comes from the Rayleigh-Ritz theorem on the eigenvalues of a symmetric matrix. For justification as to why the maximum eigenvalue of the full Gram matrix must be used here, consider a standard theorem on interlacing eigenvalues. (See for example, \cite{horn1988matrix} and \cite{hogben2007matrix}).

All told we have
\[\alpha'\left(\bm{I}_{nq} + \sigtau{1}^2\covXkkX{k}\right)^{-1}\alpha(1-nq\sigtau{0}^2\lambda_{max})\]
\[\quad\leq ~\alpha'\left(\bm{I}_{nq} + \sigtau{1}^2\covXkkX{k} + \sigtau{0}\covXkkX{j}\right)^{-1}\alpha.\]
This is equivalent to the statement proposed at (ii).

\item[(iii)] From part (i) of the current lemma, we have
\begin{eqnarray*}
Q_k &=& |\bm{I}_{nq} + \left(\bm{I}_q \otimes \bm{X}_{}D_k^{-1}\bm{X}_{}'\right)|^{-1/2}\\
&=&|\bm{I}_{nq} + \sigtau{1}^2\covXkkX{k} + \sigtau{0}^2\covXkkX{k^c}|^{-1/2}
\end{eqnarray*}
Define a matrix $G$ as follows:
\[G := \bm{I}_{nq} + \sigtau{1}^2\covXkkX{k\land t} + \sigtau{0}^2\covXkkX{k^{c}\lor t^c}.\]
By (ii) of the current lemma, we now have
\begin{align*}
&(1-\xi_n)\left(\bm{I}_{nq} + \sigtau{1}^2\covXkkX{k \land t}\right)^{-1}\\
& \leq G^{-1}\\
&\leq \left(\bm{I}_{nq} + \sigtau{1}^2\covXkkX{k \land t}\right)^{-1}.
\end{align*}

This inequality along with Condition 5 implies the following array of equalities and inequalities.
\footnotesize
\begin{align*}
&\frac{Q_k}{Q_{k \land t}}= \left|\bm{I}_{nq} + \sigtau{1}^2\covXkkX{k} + \sigtau{0}^2\covXkkX{k^{c}}\right|^{-1/2}|G|^{1/2}\\
&= \left|G + (\sigtau{1}^2 - \sigtau{0}^2)\covXkkX{k \land t^c}\right|^{-1/2}|G|^{1/2}\\
&=\left|\bm{I}_{nq} + (\sigtau{1}^2 - \sigtau{0}^2)\Xkp{k \land t^c}G^{-1}\Xk{k\land t^c}\right|^{-1/2}\\
&\leq \left|\bm{I}_{nq} + (\sigtau{1}^2 - \sigtau{0}^2)(1 - \xi_n)\right.\\
&  \quad\left.\Xkp{k \land t^c}\left(\bm{I}_{nq} + \sigtau{1}^2\covXkkX{k \land t}\right)^{-1}\Xk{k \land t^c}\right|^{-1/2}\\
&=\left|\bm{I}_{nq} + \sigtau{1}^2 \covXkkX{t} + (1-h)\sigtau{1}^2\covXkkX{k \land t^c}\right|^{-1/2}\\
& \quad \quad\left|\bm{I}_{nq} + \sigtau{1}^2\covXkkX{k \land t}\right|^{1/2}\\
&\leq \left|\bm{I}_{nq} + \sigtau{1}^2(1-h)\covXkkX{k}\right|^{-1/2}\\
& \quad\quad \left|\bm{I}_{nq} + \sigtau{1}^2\covXkkX{k \land t}\right|^{1/2}\\
&\leq (nq\sigtau{1}^2\lambda_{min}(1-h))^{-(r_k^* - r_{k \land t})/2}(1-h)^{-|k \land t|/2},
\end{align*}
\normalsize
where $(1-h) = (\tau_1^2 - \tau_0^2)\sigma_\beta^2(1- \xi_n)/\sigtau{1}^2 \to 1$ as $n \to \infty$.

Similarly, if we let \[G = \bm{I}_{nq} + \sigtau{1}^2\covXkkX{t} + \sigtau{0}^2\covXkkX{t^c},\] we can obtain by the same reasoning as above
\footnotesize
\begin{eqnarray*}
\frac{Q_{k\land t}}{Q_{t}} &=& \left|G - (\sigtau{1}^2 - \sigtau{0}^2)\covXkkX{k \land t^c}\right|^{-1/2}|G|^{1/2}\\
&\leq& \left|\bm{I}_{nq} + \sigtau{1}^2\covXkkX{k \land t}\right|^{-1/2}\\
&& \quad \quad \left|\bm{I}_{nq}+ \sigtau{1}^2\covXkkX{t}\right|^{1/2}\\
&\leq& \left|\bm{I}_{nq} + \sigtau{1}^2\covXkkX{k^c \land t}\right|^{1/2}\\
&\leq& (nq\sigtau{1}^2c)^{|k^c \land t|/2},
\end{eqnarray*}
\normalsize
where $c$ is some positive constant.
Thus it now follows that \[\frac{Q_k}{Q_{k \land t}}\cdot \frac{Q_{k \land t}}{Q_t} = \frac{Q_k}{Q_t} \leq w\lambda_{min}^{-|k^c \land t|/2}(nq\sigtau{1}^2\lambda_{min}(1-h))^{(r_k^* - r_{t})/2}\]
 for some $w > 0$.
 This completes the proof of (iii).
\end{itemize}

\section*{Proofs of Theorems in Section A.2}\label{theoremProof}

\subsection*{Prefacing lemmas}\label{prefaceLemmas}
We first establish some preliminary statements. Lemma 1 uses the posterior ratio $\text{PR}(k,t)$. However, as the Bayes factor, $\text{BF}(k,t)$, conforms better with our proof, we will use that instead. Note the following:
\[\text{PR}(k,t) = \frac{\mathbb{P}(\bm{Z} = k | \bm{Y}, \bm{X}, \sigma_\beta^2, \bm{\Sigma}_Y)}{\mathbb{P}(\bm{Z} = t | \bm{Y}, \bm{X}, \sigma_\beta^2, \bm{\Sigma}_Y)}
= \text{BF}(k,t)\cdot\frac{\mathbb{P}(\bm{Z} = k )}{\mathbb{P}(\bm{Z} = t)}.\]
However, as $\mathbb{P}(\bm{Z} = k) = \mathbb{P}(\bm{Z} = t)$ for any model $k$, we have in fact that $\text{PR}(k,t) = \text{BF}(k,t)$.

Define the following:
\normalsize
\[R_k  =  \vec{\bm{Y}}'\bigg(\fullcov-\left(\bm{\Sigma}_Y^{-1} \otimes \bm{X}_{k}\right)\mathcal{J}_k^{-1}\left(\bm{\Sigma}_Y^{-1} \otimes \bm{X}_{k}'\right)\bigg)\!\!\vec{\bm{Y}},\]
where
\[\mathcal{J}_k = \left(\sigtau{1}^{-2}\bm{I}_{nq} + \covkXXk{k}\right).\]

\tO{

Lemma 1 uses $\rtild - \widetilde{R}_t$, which approximates $R_k - R_t$. Since $R_k-R_t$ is often easier to work with, we will use it when convenient.
We now define two lemmas to assist us in establishing Theorems in Section 3.2.
\begin{lemma}\label{LemmaA1}
Let $A$ be any matrix satisfying $A \geq \bm{I}_{nq}$ and let $k$ represent a model which contains another model $j$ (i.e. $j \subset k$). We can then state the following:
\begin{itemize}
\item[(i)] $\widetilde{R}_k = \vec{\bm{Y}}\left(\fullcov +\mathcal{L}\right)^{-1}\!\vec{\bm{Y}},$
where \[\mathcal{L} = \sigtau{1}^2\left(\bm{\Sigma}_Y^{-2} \otimes \bm{X}_{k}\bm{X}_{k}'\right) + \sigtau{0}^2\left(\bm{\Sigma}_Y^{-2} \otimes \bm{X}_{j}\bm{X}_{j}'\right).\]

\item[(ii)] $\left(A + \sigtau{0}^2\covXkkX{k}\right)^{-1} - \left(A + \sigtau{1}^2\covXkkX{k}\right)^{-1}$\\
$~~~~~~~~~~~~~~~~\geq ~\left(A^{-1} - \left(A + \sigtau{1}^2\covXkkX{k}\right)^{-1}\right)(1-w_n),$ and

\item[(iii)] $(R_j - R_k)(1-w_n)(1-\xi_n)^2 \leq \widetilde{R}_j - \rtild \leq (R_j - R_k)(1-\xi_n)^{-1},$
\end{itemize}
where $w_n = o(1)$ for all allowable choices of $k$ and $j$ (i.e. $j \subset k$), and $\xi_n$ is as defined in Lemma \ref{lemma111}.
\end{lemma}

We prove each statement of Lemma \ref{LemmaA1} individually.

\begin{itemize}
\item[(i)] Borrowing from the proof of Lemma \ref{lemma111}, we once again use the Sherman-Morrison-Woodbury matrix identity to get the following:

\footnotesize

\begin{align*}
&\Xk{}\left((\bm{I}_q \otimes D_k) + \covkXXk{}\right)^{-1}\Xkp{} \\
&= \left(\bm{\Sigma}_Y^{-1} \otimes \bm{X}D_k^{-1}\bm{X}'\right)  \\
&\quad \quad- \left(\bm{\Sigma}_Y^{-1} \otimes \bm{X}D_k^{-1}\bm{X}'\right)\left(\bm{I}_{nq} +\left(\bm{\Sigma}_Y^{-1} \otimes \bm{X}D_k^{-1}\bm{X}'\right)\right)^{-1} \left(\bm{\Sigma}_Y^{-1} \otimes \bm{X}D_k^{-1}\bm{X}'\right)\\
&= \left(\bm{\Sigma}_Y^{-1} \otimes \bm{X}D_k^{-1}\bm{X}'\right)- \left(\bm{\Sigma}_Y^{-1} \otimes \bm{X}D_k^{-1}\bm{X}'\right)\left(\bm{I}_{nq} - \left(\bm{I}_{nq} + \left(\bm{\Sigma}_Y^{-1} \otimes \bm{X}D_k^{-1}\bm{X}'\right)\right)^{-1}\right)\\
&=\left(\bm{\Sigma}_Y^{-1} \otimes \bm{X}D_k^{-1}\bm{X}'\right)\left(\bm{I}_{nq} + \left(\bm{\Sigma}_Y^{-1} \otimes \bm{X}D_k^{-1}\bm{X}'\right)\right)^{-1}\\
&= \bm{I}_{nq} - \left(\bm{I}_nq + \left(\bm{\Sigma}_Y^{-1} \otimes \bm{X}D_k^{-1}\bm{X}'\right)\right)^{-1},
\end{align*}

\normalsize
which then implies
\footnotesize
\begin{align*}
&\rtild = \vec{\bm{Y}}' \fullcov\vec{\bm{Y}} \\
&\quad\quad- \vec{\bm{Y}}'\left(\bm{\Sigma}^{-1}\otimes \bm{X}\right)\left((\bm{I}_q \otimes D_k) + \covkXXk{}\right)^{-1}\left(\bm{\Sigma}^{-1}\otimes \bm{X}\right)\!\!\vec{\bm{Y}}\\
&=\vec{\bm{Y}}'\left(\fullcov + \left(\bm{\Sigma}^{-1}\otimes \bm{X}D_k^{-1}\bm{X}'\right)\right)^{-1}\!\!\vec{\bm{Y}}\\
&= \vec{Y}'\left(\fullcov + \mathcal{L}\right)^{-1}\!\!\vec{\bm{Y}}.
\end{align*}

\normalsize
This completes the proof of (i).

\item[(ii)] Taking any matrix $A \geq \bm{I}_{nq}$, we look at the left hand side (LHS) of the statement of (ii) from above.
\footnotesize
\begin{eqnarray*}
LHS &=& \left(A + \sigtau{1}^2 \covXkkX{k}\right)^{-1} - \left(A + \sigtau{0}^2\covXkkX{k}\right)^{-1}\\
&=& A^{-1}\Xk{k}\left(\sigtau{0}^{-2}\bm{I}_{|k|q} + \mathcal{M}_k\right)^{-1}\Xkp{k}A^{-1}\\
&&\quad - A^{-1}\Xk{k}\left(\sigtau{1}^{-2}\bm{I}_{|k|q} + \mathcal{M}_k\right)^{-1}\Xkp{k}A^{-1}\\
&=& A^{-1}\Xk{k}\\
&& \quad \quad \quad \times~ U\left(\left(\sigtau{0}^{-2}\bm{I}_{|k|q} + D\right)^{-1} - \left(\sigtau{1}^{-2}\bm{I}_{|k|q} + D\right)^{-1}\right)U'\\
&&\quad \quad \quad \times~\Xkp{k}A^{-1},
\end{eqnarray*}

\normalsize
where $UDU'$ is the diagonalization of \[\mathcal{M}_k:=\Xkp{k}A^{-1}\Xk{k}.\] The diagonal entries of $D$ can be denoted by $d_i$ for $1 \leq i \leq |k|q$. These are the eigenvalues of $\mathcal{M}_k$. Since $A \geq \bm{I}_{nq}$, it follows that $\bm{I}_{nq} \geq A^{-1}$ and hence $d_i \leq nq\lambda_{max}$ for each $i$.

Thus, the $i^{th}$ diagonal entry of  \[\left(\sigtau{0}^{-2}\bm{I}_{|k|q} + D\right)^{-1} - \left(\sigtau{1}^{-2}\bm{I}_{|k|q} + D\right)^{-1}\] is given by
\[\frac{1}{\sigtau{1}^{-2} + d_i}-\frac{1}{\sigtau{0}^{-2} + d_i}. \]
We then have the following relationship:
\small
\begin{eqnarray*}
\frac{1}{\sigtau{1}^{-2} + d_i}-\frac{1}{\sigtau{0}^{-2} + d_i} &=& \frac{\sigtau{1}^2 - \sigtau{0}^2}{(1 + \sigtau{1}^2d_i)(1 + \sigtau{0}^2d_i)}\\
&\geq& \frac{1-\tau_0^2/\tau_1^2}{(\sigtau{1}^{-2} + d_i)(1 + nq\sigtau{0}^2\lambda_{max})}\\
&=& \frac{1-w_n}{\sigtau{1}^{-2} + d_i},
\end{eqnarray*}
\normalsize
where $w_n = 1-\frac{1-\tau_0^2/\tau_1^2}{(1 + nq\sigtau{0}^2\lambda_{max})} \to 0$ as $n \to \infty$.
Therefore we have
\begin{eqnarray*}
LHS &\geq& A^{-1}\Xk{k}\\
&& \quad \quad \quad \times ~ U\left(\sigtau{1}^{-2}\bm{I}_{|k|q} + D\right)^{-1}U'\\
&&\quad \quad \quad \times ~{k}A^{-1}(1-w_n)\\
&=& \left(A^{-1} - \left(A + \sigtau{1}^2\covXkkX{k}\right)^{-1}\right)(1-w_n).
\end{eqnarray*}
This completes the proof of (ii).

\item[(iii)] Define matrices $A$ and $B$ as follows:
\vspace{-1em}
\footnotesize
\begin{eqnarray*}
 A&=& \bm{I}_{nq} + \sigtau{1}^2\covXkkX{j} + \sigtau{0}^2\covXkkX{k^c};\\
 B &=&\bm{I}_{nq} + \sigtau{1}^2\covXkkX{j}.
\end{eqnarray*}
We know directly from statement (ii) of Lemma \ref{lemma111} that $(1-\xi_n)B^{-1} \leq A^{-1}.$ Furthermore, by (i) of the current lemma, we have
\begin{align*}
&\widetilde{R}_j - \rtild\\
 &= \Upsilon'\left(A + \sigtau{0}^2\covXkkX{k \land j^c}\right)^{-1}\Upsilon\\
 &\quad \quad-\Upsilon'\left(A + \sigtau{1}^2\covXkkX{k \land j^c}\right)^{-1}\Upsilon\\
&\leq \Upsilon' A^{-1}\Upsilon- \Upsilon'\left(A + \sigtau{1}^2\covXkkX{k \land j^c}\right)^{-1}\Upsilon\\
&= \Upsilon' A^{-1}\Xk{k \land j^c}\\
&\quad\quad\times\left(\sigtau{1}^{-2}\bm{I}_{|k \land j^c|q} + \Xkp{k \land j^c}A^{-1}\Xk{k \land j^c}\right)^{-1}\\
&\quad\quad\times\Xkp{k \land j^c} A^{-1}\Upsilon\\
\end{align*}
\begin{align*}
&\leq\Upsilon' B^{-1}\Xk{k \land j^c}\\
&\quad\quad\times\left(\sigtau{1}^{-2}\bm{I}_{|k \land j^c|q} + \Xkp{k \land j^c}B^{-1}(1-\xi_n)\Xk{k \land j^c}\right)^{-1}\\
&\quad\quad\times\Xkp{k \land j^c}B^{-1}\Upsilon\\
&= \Upsilon' B^{-1}\Xk{k \land j^c}\\
&\quad\quad\times\left(\sigtau{1}^{-2}(1-\xi_n)^{-1}\bm{I}_{|k \land j^c|q} + \Xkp{k \land j^c}B^{-1}\Xk{k \land j^c}\right)^{-1}\\
&\quad\quad\times\Xkp{k \land j^c}B^{-1}\Upsilon(1-\xi_n)^{-1}\\
&=\left(\Upsilon' B^{-1}\Upsilon-\Upsilon'\left(B + \sigtau{1}^2(1-\xi_n)\covXkkX{k \land j^c}\right)^{-1}\Upsilon\right)(1-\xi_n)^{-1}\\
&\leq (R_j - R_k)(1-\xi_n)^{-1},
\end{align*}
\normalsize
where $\Upsilon = \sqrtcov\vec{\bm{Y}}$.

Now from (ii) of the current lemma, we have
 \small
\begin{align*}
&\widetilde{R}_j - \rtild \\
&\geq \left(\Upsilon' A^{-1}\Upsilon- \Upsilon'\left(A + \sigtau{1}^2\covXkkX{k \land j^c}\right)^{-1}\Upsilon\right)(1-w_n)\\
&=\Upsilon' A^{-1}\Xk{k \land j^c}\\
&\quad\quad\times\left(\sigtau{1}^{-2}\bm{I}_{|k \land j^c|q} + \Xkp{k \land j^c}A^{-1}\Xk{k \land j^c}\right)^{-1}\\
&\quad\quad\times\Xkp{k \land j^c}A^{-1}\Upsilon(1-w_n)\\
&\geq\Upsilon' B^{-1}\Xk{k \land j^c}\\
&\quad\quad\times\left(\sigtau{1}^{-2}\bm{I}_{|k \land j^c|q} + \Xkp{k \land j^c}B^{-1}\Xk{k \land j^c}\right)^{-1}\\
&\quad\quad\times\Xkp{k \land j^c}B^{-1}\Upsilon(1-w_n)(1-\xi_n)^2\\
&=\left(\Upsilon' B^{-1}\Upsilon - \Upsilon'\left(B + \sigtau{1}^2(1-\xi_n)\covXkkX{k \land j^c}\right)^{-1} \Upsilon\right)\\
&\quad\quad\quad\quad \times (1-w_n)(1-\xi_n)^{2}\\
&= (R_j - R_k)(1-w_n)(1-\xi_n)^{2}.
\end{align*}

\normalsize
This completes the proof of (iii).
\end{itemize}

}

\normalsize

\tO{
We will denote by $R_k^*$ the residual sum of squares for the ordinary least squares estimate of $\bm{\beta}$ under the assumption of model $k$. That is,
$R_k^* = \vec{Y}'\left(\bm{I}_{nq} - P_k\right)\vec{Y},$ where $P_k$ is the standard projection matrix given by
\[P_k = \left(\bm{I}_q \otimes X_k\right)\left(\bm{I}_q \otimes X_k'X_k\right)^{-1}\left(\bm{I}_q \otimes X_k'\right).\]
We can now state the following about $\rtild$, $R_k$, and $R_k^*$, for any model $k$.

By definition, $\rtild \leq R_k$ (since $\rtild$ is the residual sum of squares for the minimizer of the general least squares estimate of $\bm{\beta}$; this is a special case of the Gauss-Markov theorem. See \cite{izenman2008MV}). Also, $R_k^* \leq R_k$ since if $X_k = U_k\Lambda_k V_k'$ is the singular value decomposition of $X_k$, then
\begin{eqnarray*}
P_k&=& \left(\bm{I}_q \otimes U_k\Lambda_k V_k'\right)\left(\left[\bm{I}_q\otimes V_k\Lambda_k' U_k'\right]\left[\bm{I}_q\otimes U_k\Lambda_k V_k'\right] \right)^{-1}\left(\bm{I}_q \otimes V_k\Lambda_k' U_k'\right)\\
&=&\left(\bm{I}_q \otimes U_k\Lambda_k V_k'\right)\left(\bm{I}_q\otimes V_k\Lambda_k' U_k'U_k\Lambda_k V_k'\right)^{-1}\left(\bm{I}_q \otimes V_k\Lambda_k' U_k'\right)\\
&=&\left(\bm{I}_q \otimes U_k\Lambda_k V_k'\right)\left(\bm{I}_q\otimes V_k(\Lambda_k'\Lambda_k)^{-1} V_k'\right)\left(\bm{I}_q \otimes V_k\Lambda_k' U_k'\right)\\
&=& \bm{I}_q \otimes \left[U_k\Lambda_k V_k' V_k (\Lambda_k'\Lambda_k)^{-1}V_k'V_k\Lambda_k'U_k'\right]\\
&=& \bm{I}_q \otimes \left[U_k\Lambda_k (\Lambda_k'\Lambda_k)^{-1}\Lambda_k'U_k'\right].
\end{eqnarray*}

However, we also have
\begin{align*}
&\Xk{k}\left(\sigtau{1}^{-2}\bm{I}_{n|k|} + \covkXXk{k}\right)^{-1}\Xkp{k}\\
&= \left(\bm{\Sigma}_Y^{-1} \otimes U_k \Lambda_k\right)\left(\sigtau{1}^{-2}\bm{I}_{n|k|} + \bm{\Sigma}_Y^{-1} \otimes \Lambda_k'\Lambda_k\right)^{-1}\left(\bm{\Sigma}_Y^{-1} \otimes \Lambda_k' U_k'\right)\\
&\geq \bm{I}_q \otimes \left[U_k\Lambda_k (\Lambda_k'\Lambda_k)^{-1}\Lambda_k'U_k'\right].
\end{align*}
It thus follows that $R_k^* \leq R_k$.
\normalsize

\textcolor{black}{We want to add a comment on the matrix $\bm{\Sigma}_{0,Y}$. Since each column $Y_{(j)}$ of $\bm{Y}$ can be scaled such that $\text{Var}(Y_{(j)}) = \sigma_{0,Y}^2$, we can assume that the diagonal of $\bm{\Sigma}_{0,Y}$ consists of the common variance $\sigma_{0,Y}^2$ repeated $q$ times. Per the comments in Chapter 6 of \cite{izenman2008MV} and Chapter 15 of \cite{krzanowski2000}, the residual sum of squares for any column $\beta_{(j)}$ of $\bm{\beta}$ (where $\beta_{(j)}$ is of course the coefficient vector associated with $Y_{(j)}$) estimated by OLS is given by}
\[\text{RSS}_j^* = Y_{(j)}'(\bm{I}_n - X_k(X_k'X_k)^{-1}X_k')Y_{(j)}.\]
Moreover, the residual vector $\hat{e}_{(j)}$ associated with $\text{RSS}_j^*$ has distribution
\[\hat{e}_{(j)} \sim N_q\left(\bm{0}, \sigma_{0,Y}^2(\bm{I}_n -  X_k(X_k'X_k)^{-1}X_k')\right).\]
}

\tO{
Since $\frac{\text{RSS}_j^*}{\sigma_{0,Y}^2} \sim \chi^2_{n - |k|}$ and $R_k^* = \sum_{j = 1}^q \text{RSS}_j^*$, then $R_k^*/\sigma_{0,Y}^2$ has the same distribution as the sum of the $\text{RSS}_j^*/\sigma_{0,Y}^2$. The sum of $q$-many $\chi^2_{n - |k|}$ random variables has distribution $\chi^2_{q(n - |k|)}$. Hence we have \[\frac{R_k^*}{\sigma_{0,Y}^2} \sim\chi^2_{q(n - |k|)}. \]


We now establish a lemma that bounds the difference between $R_t$ and $R_t^*$.

}
\tO{
\begin{lemma}\label{LemmaA2}
Let $\{g_n\}_{n = 1}^{\infty}$ be any real sequence diverging to $\infty$. Take $\epsilon > 0$. We then have the following:
\begin{itemize}
\item[(i)]$\mathbb{P}(R_t - R_t^* > g_n) \leq \text{exp}\left\lbrace-cnq\sigtau{1}^2g_n\right\rbrace$;
\item[(ii)]$ \mathbb{P}\left(\left|\frac{R_t^*}{nq\sigma_{0,Y}^2} -1\right| > \epsilon\right)\leq \text{exp}\left\lbrace-cnq\right\rbrace$,
\end{itemize}
where in both items $c>0$ is some constant.
\end{lemma}

\begin{itemize}
\item[(i)] By (6.13) and (6.14) of \cite{izenman2008MV}, and aided once again by the Sherman-Morrison-Woodbury identity, we have
\small
\begin{eqnarray*}
0 &\leq& R_t - R_t^*\\
&=& \vec{Y}'\left(\bm{\Sigma}_Y^{-1} \otimes \bm{X}_{t}\right)\\
&& \quad \quad \quad \times \left(\covkXXk{t}^{-1} - \left(\sigtau{1}^{-2}\bm{I}_{|t|q} +\covkXXk{t}\right)^{-1} \right)\\
&& \quad \quad \quad \times\left(\bm{\Sigma}_Y^{-1} \otimes \bm{X}_{t}'\right)\vec{Y}\\
&=&\vec{Y}'\left(\bm{I}_q \otimes \bm{X}_{t}(\bm{X}_t'\bm{X}_t)^{-1}\right)\\
&& \quad \quad \quad \times\left(\sigtau{1}^{2}\bm{I}_{|t|q}+\covkXXk{t}^{-1}\right)^{-1}\\
&& \quad \quad \quad \times\left(\bm{I}_q \otimes (\bm{X}_t'\bm{X}_t)^{-1}\bm{X}_{t}'\right)\vec{Y}\\
&=&(nq\sigtau{1}^2)^{-1}\vec{Y}'M\vec{Y},
\end{eqnarray*}
\normalsize
where $M = n \left(\bm{I}_q \otimes \bm{X}_{t}(\bm{X}_t'\bm{X}_t)^{-2}\bm{X}_{t}'\right)$. Note that $M$ has rank $|t|$ and the eigenvalues of $M$ are bounded. We thus can conclude
\begin{eqnarray*}
\mathbb{P}(R_t - R_t^* > g_n) &\leq& \mathbb{P}(\vec{Y}'M\vec{Y} > nq\sigtau{1}^2g_n)\\
&\leq& \text{exp}\left\{-cnq\sigtau{1}^2g_n\right\}.
\end{eqnarray*}
This establishes (i).

\item[(ii)] As previously discussed, $R_t^*/\sigma_{0,Y}^2$ follows a $\chi^2_{q(n - |k|)}$ distribution. By Lemma 1 of \cite{laurentmassart2000}, we have
\[\mathbb{P}\left(|R_t^*/\sigma_{0,Y}^2 - q(n-|t|)| \geq 2q(n-|t|)(\sqrt{x} + 2x)\right) \leq 2 \:\text{exp}\{-q(n-|t|)x\}.\]
From this we then obtain
\[\mathbb{P}(|R_t^*/(nq\sigma_{0,Y}^2)-1| > \epsilon) \leq \text{exp}\{-cnq\}.\]
This was the statement of (ii), thus completing the proof of Lemma \ref{LemmaA2}.
\end{itemize}

}

\tO{
\subsection*{Proof of Theorem 1}
From here we begin the main body of the proof of Theorem 1. As always, the true model will be denoted by $t$. The set of incorrect models (i.e. those not equal to $t$) can be divided into four subsets $(M_1, M_2, M_3, M_4)$ defined as follows:
\begin{itemize}
\item \textbf{Uselessly large models:} Define $M_1:= \{k \mid r_k > m_n \}.$ $M_1$ consists of all models $k$ such that the dimension of $k$ (given by $r_k$) exceeds $m_n$.
\item \textbf{Overfitted models:} Define $M_2:= \{k \mid t \subset k, r_k \leq m_n \}.$ $M_2$ consists of all models $k$ of dimension not exceeding $m_n$ such that $k$ contains the true model, plus one or more inactive covariates.
\item \textbf{Large models not containing $t$:} Define $M_3:= \{k \mid t \not\subset k, K|t| < r_k \leq m_n \}.$ $M_3$ consists of all models $k$ of dimension greater than $K|t|$, yet not exceeding $m_n$, such that $k$ is missing one or more active covariates. Here $K$ is as discussed in Condition 4.
\item \textbf{Underfitted models:} Define $M_4:= \{k \mid t \not\subset k, r_k \leq K|t| \}.$ $M_4$ consists of all models $k$ of dimension not exceeding $K|t|$ such that $k$ is missing one or more active covariates.
\end{itemize}

We will prove that $\sum_{k \in M_u}\text{BF}(k,t) \xrightarrow{\mathbb{P}}0$ for each of $u = 1,2,3,4$.
\subsubsection{The models of $M_1$} Note that in the event that $p \preceq n/\log{n}$, $M_1$ is the empty set. Take any $s > 0$. We then have:
\begin{eqnarray*}
\mathbb{P}\left(\bigcup_{k \in M_1}\left\{\widetilde{R}_t - \rtild > nq(1+2s)\sigma_{0,Y}^2\right\}\right)&\leq&\mathbb{P}\left(\widetilde{R}_t > nq(1+2s)\sigma_{0,Y}^2\right)\\
&\leq&\mathbb{P}\left(R_t > nq(1+2s)\sigma_{0,Y}^2\right)\\
&\leq& \mathbb{P}\left(R_t^* > nq(1+s)\sigma_{0,Y}^2\right)\\
&&\quad \quad\quad\quad + \mathbb{P}\left(R_t - R_t^* > nqs\sigma_{0,Y}^2\right)\\
&\leq& 2\: \text{exp}\{-cnq\},
\end{eqnarray*}
due to the results of Lemma \ref{LemmaA2}.(ii).

Now consider the term $nq\sigtau{1}^2\lambda_{min}(1-h)$, where $h$ is as defined in the proof of Lemma \ref{lemma111}. Conditions 2 and 5 imply
\begin{equation}
\max\left(p^{2 + 2\delta}, nq\right)\preceq nq\sigtau{1}^2\lambda_{min}(1-h) \preceq \max\left(p^{2 + 3\delta}, nq\right).\label{(TWO)}
\end{equation}
Examining the high probability event $\{\widetilde{R}_t - \rtild \leq nq(1+2s)\sigma_{0,Y}^2\}$, Lemma 1 and (\ref{(TWO)}) give the following:
\begin{eqnarray*}
\sum_{k \in M_1}\text{BF}(k,t) &\preceq& \sum_{k \in M_1}p^{-(1+\delta)(m_n-|t|)}s_n^{|k|-|t|}\lambda_{min}^{-|t|/2}e^{nq(1+2s)/2}\\
&\preceq& \sum_{k \in M_1}e^{-nq(1+\delta)/(2+\delta)}s_n^{|k|-|t|}\lambda_{min}^{-|t|/2}e^{nq(1+2s)/2}.
\end{eqnarray*}
This follows from the fact that for any $k \in M_1$, $r_k^* = m_n > nq/\log(p^{2 + \psi}) \geq nq/\log(p^{2+\delta})$. In turn, whenever $s$ satisfies \[1 + 2s < 2(1+\delta)/(2+\delta); \quad \text{i.e. } s < \delta/2(2+\delta),\] then Condition 5 and the assumption that $s_n \sim p^{-1}$ now imply the following:
\begin{eqnarray*}
\sum_{k \in M_1}\text{BF}(k,t) &\preceq& e^{-nq(1+\delta)/(2+\delta)}e^{nq(1+2s)/2}p^{\kappa|t|}\sum_{k \in M_1} s_n^{|k| - |t|}\\
&\preceq& e^{-nq(1+\delta)/(2+\delta)}e^{nq(1+2s)/2}p^{c|t|}\sum_{|k| = m_n}^{p}\binom{p}{|k|}s_n^{|k|}\\
&\preceq& e^{-nq(1+\delta)/(2+\delta)}e^{nq(1+2s)/2}p^{c|t|}(1+s_n)^p\\
&\preceq& e^{-wnq} \to 0 ~~\text{as}~~ n \to \infty,
\end{eqnarray*}
where $w$ is a positive constant.

All told, we thus have \[\sum_{k \in M_1}\text{BF}(k,t) \xrightarrow{\mathbb{P}}0.\]
}\tO{\subsubsection{The models of $M_2$}\label{A.1.4} Our aim now is to uniformly bound $\widetilde{R}_t- \rtild$ for any model $ k \in M_2$ using deviation inequalities for quadratic forms $\varepsilon' P_{k \land t^c}\varepsilon$, where $\varepsilon$ is some real vector satisfying the array of inequalities and equalities below and $P_{k \land t^c}$ is the usual projection matrix for the model $k \land t^c$. We will use the same deviation inequalities involving the $\chi^2$ distribution as established in \cite{NNNXH:Addendum}.

Take any $k \in M_2$. We have the following:
\begin{eqnarray*}
R_t^* - R_k^* &=& \lVert (P_k - P_t)\vec{Y}\rVert_2^2\\
&\leq& \left(\vertiii{(P_k - P_t)\bm{X}_{t^c}\bm{\beta}_{t^c}} + \lVert(P_k - P_t)\varepsilon\rVert_2\right)^2\\
&\leq& \left(\vertiii{\bm{X}_{t^c}\bm{\beta}_{t^c}} + \sqrt{\varepsilon'P_{k \land t^c}\varepsilon}\right)^2\\
&=&\left(b_0 + \sqrt{\varepsilon'P_{k \land t^c}\varepsilon}\right)^2,
\end{eqnarray*}
where $b_0 = \vertiii{\bm{X}_{t^c}\bm{\beta}_{t^c}} = \mathcal{O}(1)$ as per Condition 3. Note here that we have used a slight abuse of notation in regards to the projection matrices $P_k$ and $P_t$: Where convenient, we take the projection matrices to be formed with or without the inclusion of the Kronecker product.
Ultimately, this is merely a case of looking at the same object via two different representations.
Note that as $\varepsilon'P_{k \land t^c}\varepsilon$ is a quadratic form and $P_{k \land t^c}$ has rank $r_k - r_t$, then $\varepsilon'P_{k \land t^c}\varepsilon$ follows a $\chi^2$ distribution with $r_k - r_t$ degrees of freedom.
(Remember that $t \subset k$, so that $r_k > r_t$). It follows that for any $x > 0$ and for some $\sqrt{2/3} < w' < 1$, we have
\begin{align}
\begin{split}\label{(FOUR)}
&\mathbb{P}\left(R_t^* - R_k^* > \sigma_{0,Y}^2(2 + 3x)(r_k - r_t)\log p\right)\\
&\leq \mathbb{P}\left(\varepsilon'P_{k \land t^c}\varepsilon > \sigma_{0,Y}^2(2 + 3w'x) (r_k-r_t)\log p\right)\\
&\leq \mathbb{P}\left(\chi^2_{r_k - r_t} - (r_k-r_t) > (2 + 3(w')^2x)(r_k-r_t)\log p\right)\\
&\leq c\:\text{exp}\{-(1+x)(r_k-r_t)\log p\}\\
&=cp^{-(1+x)(r_k-r_t)}.
\end{split}
\end{align}

\normalsize
Take positive $s$ satisfying $s \leq \delta/8$.
Also take a sequence $\{w_n\}_{n= 1}^{\infty}$ such that $w_n = o(1)$ when treated as a function of $n$.
Now define the event $A(k)$ for any model $k \in M_2$:
\begin{eqnarray*}
A(k) &:=& \left\{\widetilde{R}_t - \rtild > 2\sigma_{0,Y}^2(1 + 4s)(r_k-r_t)(1-w_n)\log p\right\}\\
&\subset& \left\{R_t - R_k > 2\sigma_{0,Y}^2(1 + 4s)(r_k-r_t)(1-w_n)(1-\xi_n)\log p\right\}\\
&\subset& \left\{R_t - R_k > 2\sigma_{0,Y}^2(1 + 2s)(r_k-r_t)\log p\right\},
\end{eqnarray*}
 where the subset claims are a direct result of Lemma \ref{LemmaA1}.(iii). Let $d > r_t$ be a fixed dimension. Define the event
 \[U(d):=\bigcup_{\{k: r_k=d\}}A(k).\]
Since we know that $R_k \geq R_k^*$, we now have
\begin{align}
\begin{split}\label{(FIVE)}
\mathbb{P}\left(U(d)\right) &\leq \mathbb{P}\left(\cup_{\{k: r_k=d\}}\left\{R_t - R_k^* > 2\sigma_{0,Y}^2(1 + 2s)(r_k-r_t)\log p\right\}\right)\\
&\leq\mathbb{P}\left(\cup_{\{k: r_k=d\}}\left\{R_t^* - R_k^* > \sigma_{0,Y}^2(2 + 3s)(d-r_t)\log p\right\}\right)\\
&\quad\quad\quad+ ~\mathbb{P}\left(R_t- R_t^* > s\sigma_{0,Y}^2(d-r_t)\log p\right).
\end{split}
\end{align}

The event $\left\{R_t^* - R_k^* > \sigma_{0,Y}^2(2 + 3s)(d-r_t)\log p\right\}$ will depend entirely on $P_{k \land t^c}$. It follows that the union of all such events can thus be indexed by all possible values of $k \land t^c$. There are at most $p^{d-r_t}$ such projections, since there can be at most $p^r$ subspaces of rank $r$, and any projection matrix $P_{k \land t^c}$ corresponds to a rank $d-r_t$ subspace.

Then (\ref{(FOUR)}) and (\ref{(FIVE)}), along with Lemma \ref{LemmaA2}.(i) allow us to establish
\begin{align}
\begin{split}\label{(SIX)}
\mathbb{P}\left(U(d)\right) &\leq cp^{-(1+s)(d-r_t)}p^{(d-r_t)} + \text{exp}\{-cnq\log p\}\\
&\leq 2cp^{-s(d-r_t)}.
\end{split}
\end{align}
We now can examine the union of all events $U(d)$ with $d > r_t$:
\begin{align*}
\mathbb{P}\left(\cup_{\{d > r_t\}}U(d)\right) &\leq \sum_{\{d > r_t\}}\mathbb{P}\left(U(d)\right)\\
&\leq 2c \sum_{d > r_t}p^{-s(d-r_t)}\\
&\leq 2cp^{-s}(1-p^{-s})^{-1}\\
&=\frac{2c}{p^s -1} \to 0 \quad \text{as}\quad n \to \infty.
\end{align*}

We now can restrict our attention to the high probability event $\cap_{\{d > r_t\}}U(d)^c.$ Lemma 1 and (\ref{(TWO)}) give us the following:
\begin{align*}
\sum_{k \in M_2}\text{BF}(k,t) &\preceq \sum_{k \in M_2} (nq\sigtau{1}^2\lambda_{min}(1-h))^{-(r_k-r_t)}s_n^{|k|-|t|}\text{exp}\{\rtild - \widetilde{R}_t\}\\
&\preceq \sum_{k \in M_2} \max\left(p^{1+\delta}, \sqrt{nq}\right)^{-(r_k-r_t)}s_n^{|k|-|t|}p^{(1+4s)(r_k-r_t)}\\
&\preceq \sum_{k \in M_2} \max\left(p^{\delta-4s}, \sqrt{nq}p^{-1-4s}\right)^{-(r_k-r_t)}s_n^{|k|-|t|}\\
&\preceq \min\left(p^{-\delta/2}, \frac{p^{1+\delta/2}}{\sqrt{nq}}\right)\sum_{|k| = |t| + 1}^{p}\binom{p}{|k| - |t|}s_n^{|k|-|t|}\\
&\sim \psi_n\to 0 \quad \text{as} \quad n \to \infty,
\end{align*}
where $\psi_n = \min\left(p^{-\delta/2}, \frac{p^{1+\delta/2}}{\sqrt{nq}}\right) \to 0$ as $ n \to \infty$. Note here that $r_k^* = r_k$ since $r_k < m_n$ was assumed for all $k \in M_2$. Furthermore, $(1+s_n)^{p} \sim 1$ since $s_n\sim p^{-1}$. We thus can finally conclude
\[\sum_{k \in M_2}\text{BF}(k,t) \xrightarrow{\mathbb{P}}0.\]
}

\tO{
\subsubsection*{The models of $M_3$}
Similar to the definition of $A(k)$ in the previous subsection \ref{A.1.4}, we can define for any model $k$ the event $B(k)$ as follows:
\begin{align*}
B(k)&:= \left\lbrace\widetilde{R}_t - \rtild > 2\sigma_{0,Y}^2(1+4s)(r_k-r_t)(1-w_n)\log p\right\rbrace\\
&\subset \left\lbrace\widetilde{R}_t - \widetilde{R}_{k \lor t} > 2\sigma_{0,Y}^2(1+4s)(r_k-r_t)(1-w_n)\log p\right\rbrace\\
&\subset \left\lbrace{R}_t - R_{k \lor t} > 2\sigma_{0,Y}^2(1+2s)(r_k-r_t)\log p\right\rbrace.
\end{align*}

We define the union of such events below:
\[V(d) := \bigcup_{\{k:r_k = d\}} B(k).\]
Similar to (\ref{(SIX)}), if $d > K|t|$ and $s = \delta/8$, we have
\begin{align*}
\mathbb{P}\left(V(d)\right) &\leq \mathbb{P}\left(\cup_{\{k:r_k = d\}} \left\lbrace{R}_t - R_{k \lor t} > 2\sigma_{0,Y}^2(1+2s)(r_k-r_t)\log p\right\rbrace\right)\\
&\leq cp^{-(1 + s)(d-r_t) + d}\\
&\leq cp^{-(1+w)d + d}\\
&= cp^{-wd},\end{align*}
where the inequality $(1+s)(d-r_t) > (1+w)r_k$ holds for some $w >0$ (since $(d-r_t)/d > (K-1)/K > 1/(1+\delta/8)$), which implies $(1+s)(d-r_t) > r_k$. Thus
\[\mathbb{P}\left(\cup_{\{d >K|t|\}}V(d)\right) \leq p^{-wK|t|} \to 0 \quad \text{as} \quad n \to \infty.\]
Now restricting our attention to the event $\cap_{\{d > r_t\}}V(d)^c$, with probability at least \[1-\text{exp}\{-wK|t|\log p\} \to 1,\] we have the following:
\begin{align*}
&\sum_{k \in M_3} \text{BF}(k,t)\\ &\preceq \sum_{k \in M_3} \max\left(p^{1+\delta}, \sqrt{nq}\right)^{-(r_k-r_t)}(\lambda_{min})^{-|k^c\land t|/2} s_n^{|k|-|t|}p^{(1+4s)(r_k-r_t)}\\
&\preceq \sum_{k \in M_3} \max\left(p^{\delta-4s}, \sqrt{nq}p^{-1-4s}\right)^{-(r_k-r_t)}(\lambda_{min})^{-|t|/2}s_n^{|k|-|t|}\\
&\preceq \min\left(p^{-\delta/2}, \frac{p^{1+2s}}{\sqrt{nq}}\right)^{(K-1)r_t+1}p^{\kappa|t|/2}\sum_{k \in M_3}s_n^{|k|-|t|}\\
&\preceq \psi_n^{(K-1)r_t+1}p^{\delta(K-1)|t|/4}(1 + s_n)^{p}\\
&\sim \psi_n^{(K-1)|t|/2} \rightarrow 0 \text{ as } n \to \infty.
\end{align*}
Note that $\kappa$ is as defined in Condition 5. We can now conclude that
\[\sum_{k \in M_3}\text{BF}(k,t) \xrightarrow{\mathbb{P}}0.\]

}
\tO{\subsubsection*{The models of $M_4$} Take $c' \in (0,1)$. We prove that for such $c'$,
\[\mathbb{P}\left(\bigcup_{k \in M_4}\left\lbrace\rtild - \widetilde{R}_t < \Delta_n(1-c')\right\rbrace\right) \to 0 \quad\text{as}\quad n \to \infty.\]
Here $\Delta_n$ is as defined with Condition 4. As before, let $\varepsilon$ be some real $nq$-vector such that the following array of equalities and inequalities is satisfied. (Condition 3 guarantees that such an $\varepsilon$ exists). We have the following;
\begin{align*}
R_k^* - R_{k \lor t}^* &= \left(\vertiii{(P_{k \lor t} -P_k)\bm{Y}}\right)^2\\
&= \lVert(P_{k \lor t} -P_k)\left(\bm{I}_q \otimes \bm{X}_t\right)\vec{\beta_t}  + (P_{k \lor t} -P_k)\varepsilon\rVert_2^2\\
&\geq \bigg(\vertiii{(P_{k \lor t} -P_k)\bm{X}_t\bm{\beta}_t} - \lVert(P_{k \lor t} -P_k)\varepsilon\rVert_2\bigg)^2,
\end{align*}
where the last line of  the statement follows from a simple application of the triangle inequality. (In short, $|a + b - b| \leq |a|$ for any $a,b$).
By Condition 4,
\[\vertiii{(P_{k \lor t} -P_k)\bm{X}_t\bm{\beta}_t} = \vertiii{(I -P_k)\bm{X}_t\bm{\beta}_t} \geq \sqrt{\Delta_n}.\] Thus for any $w \in (0,1)$, we have
\begin{align*}
&\mathbb{P}\left(\cup_{k \in M_4}\left\{R_k^* - R_{k \lor t}^* < (1-w)^2\Delta_n\right\}\right)\\
 &\leq \mathbb{P}\left(\cup_{k \in M_4}\left\{\lVert(P_{k \lor t} -P_k)\varepsilon\rVert_2 > w\sqrt{\Delta_n}\right\}\right)\\
&\leq \mathbb{P}\left(\lVert P_t\varepsilon\rVert_2 > w \sqrt{\Delta_n}\right)\\
&\leq \text{exp}\{-c \Delta_n\}.
\end{align*}

Since $R_k \geq R_k^*$, it follows that for any $w' \in (0,1)$ we have the following:
\begin{align}
\begin{split}\label{(ELEVEN)}
&\mathbb{P}\left(\cup_{k \in M_4}\left\{R_k - R_{k \lor t} < \Delta_n(1-w')\right\}\right)\\
&\quad\quad\leq \mathbb{P}\left(\cup_{k \in M_4}\left\{R_k^* - R_{k \lor t}^* < \Delta_n(1-w'/2)\right\}\right)\\
&\quad\quad\quad\quad + \mathbb{P}\left(\cup_{k \in M_4}\left\{R_{k \lor t}^* - R_{k \lor t} < \Delta_n(w'/2)\right\}\right)\\
&\quad\quad \leq 2\: \text{exp}\{-c\Delta_n\}\to 0 \quad\text{as}\quad n \to \infty.
\end{split}
\end{align}

This last inequality can be established in a similar manner as to that done in Section \ref{prefaceLemmas} for the exponential tails bounding the difference between $R_t$ and $R_t^*$.
To wit, let $X_{k \lor t} = U\Lambda V'$ be the singular value decomposition of $X_{k \lor t}$. Here $U$ is an $n \times n$ matrix, $\Lambda$ is an $n \times |k \lor t|$ matrix, and $V$ is a $|k \lor t| \times |k \lor t|$ matrix. It follows that
\[P_{k \lor t} = \bm{I}_q \otimes U\Lambda(\Lambda'\Lambda)^{-1}\Lambda'U'.\]
Note that $P_{k \lor t}$ is the projection matrix onto the column space of $X_{k \lor t}$. Hence we have
\begin{align*}
0 &\leq R_{k \lor t}^* - R_{k\lor t}\\
&= \vec{Y}'(\bm{\Sigma}_Y^{-1} \otimes U\Lambda)\\
&\quad \quad\quad\times\left(\left(\sigtau{1}^{-2}\bm{I}_{|k \lor t|q} + (\bm{\Sigma}_Y^{-1} \otimes \Lambda'\Lambda)\right)^{-1} -\bm{I}_{|k \lor t|q}\right)\\
&\quad\quad\quad\times (\bm{\Sigma}_Y^{-1} \otimes \Lambda' U') \vec{Y}\\
&= \sigtau{1}^{-2}\vec{Y}'\left(\bm{\Sigma}_Y^{-1} \otimes U \Lambda\right)\\
&\quad \quad\quad\times\left(\sigtau{1}^{-2}\bm{I}_{n|k \lor t|} + \bm{\Sigma}_Y^{-1} \otimes \Lambda'\Lambda\right)^{-1}\\
&\quad \quad\quad\times\left(\bm{\Sigma}_Y^{-1} \otimes \Lambda' U'\right)\vec{Y}\\
&\leq (nq\sigtau{1}^2\lambda_{min})^{-1} \vec{Y}'(\bm{I}_q \otimes U\Lambda(\Lambda'\Lambda)^{-1}\Lambda'U')\vec{Y}.
\end{align*}
\normalsize

Since $U\Lambda(\Lambda'\Lambda)^{-1/2}$ is a matrix of rank not exceeding $(K+1)|t|$ (consider Condition 3), we now have
\begin{align*}
&\mathbb{P}\left(\cup_{k \in M_4}\left\{R_{k \lor t}^* - R_{k \lor t} < -\Delta_nw'/2\right\}\right) \\&\preceq \text{exp}\left\{-wnq\sigtau{1}^2\lambda_{min}\Delta_n \right\}p^{(K+1)|t|)}\\
&\preceq \text{exp}\left\{-p^{2+\delta}\Delta_n + (K+1)|t|\log p\right\}\\
&\preceq \text{exp}\left\{-c\Delta_n\right\}.
\end{align*}

By combining (\ref{(ELEVEN)}), Lemma \ref{LemmaA1}, and Lemma \ref{LemmaA2}, and taking $0 \leq c' = 2w' < 1$, we now have
\begin{align*}
&\mathbb{P}\left(\cup_{k \in M_4}\left\{\rtild - \widetilde{R}_t < \Delta_n(1-c')\right\}\right)\\
&\quad\quad \leq \mathbb{P}\left(\cup_{k \in M_4}\left\{\rtild -\widetilde{R}_{k \lor t} < \Delta_n(1-2w')\right\}\right)\\
&\quad \quad\quad\quad\quad + \mathbb{P}\left(\cup_{k \in M_4}\left\{\widetilde{R}_ {k \lor t} - \widetilde{R}_t < -\Delta_nw'\right\}\right)\\
&\quad\quad \leq \mathbb{P}\left(\cup_{k \in M_4}\left\{R_k -{R}_{k \lor t} < \Delta_n(1-w')\right\}\right)\\
&\quad \quad\quad\quad\quad + \mathbb{P}\left(\cup_{k \in M_4}\left\{R_t - {R}_{k \lor t} > \Delta_n(w')^2\right\}\right)\\
&\quad\quad\leq \text{exp}\{-c\Delta_n\} + \mathbb{P}\left(\cup_{k \in M_4}\left\{R_t^* - {R}_{k \lor t}^* > (w')^2\Delta_n\right\}\right)\\
&\quad \quad\quad\quad\quad +\mathbb{P}\left(R_t - R_t^* > (w')^2\Delta_n\right)\\
\end{align*}
\begin{align*}
&\quad\quad\leq  \mathbb{P}\left(\chi_{K|t|}^2 > (w')^2\Delta_n\right) + 2 \text{exp}\{-c\:\Delta_n\}\\
&\quad\quad\leq 3 \:\text{exp}\{-c\:\Delta_n\}\to 0\quad\text{for all $k \in M_4$}\text{ as } n \to \infty.
\end{align*}

Define the event $C_n$ as follows:
\[C_n := \left\{\rtild - \widetilde{R}_t \geq \Delta_n(1-c'), \forall k \in M_4\right\}.\]
Thus $C_n$ is the event that $\rtild - \widetilde{R}_t$ is greater than or equal to $\Delta_n(1-c')$ for every $k \in M_4$. This is the converse event on the first line of the above array of inequalities. As shown above, $\mathbb{P}(C_n) \geq 1-3\:\text{exp}\{-c\Delta_n\}$. We now employ Conditions 2 and 4 to get
\begin{align*}
&\sum_{k \in M_4}\text{BF}(k,t) \preceq \sum_{k \in M_4}(nq\sigtau{1}^2\lambda_{min})^{|t|/2}\lambda_{min}^{-|t|/2}s_n^{|k|-|t|}\text{exp}\left\{-\frac{1}{2}(\rtild - \widetilde{R}_t)\right\}\\
&\quad\quad \preceq \sum_{k \in M_4}\max\left(p^{2+3\delta}, nq\right)^{|t|/2}p^{\delta|t|/2}s_n^{|k|-|t|}\text{exp}\left\{-\Delta_n(1-c')/2\right\}\\
&\quad\quad\preceq \text{exp}\left\{-\frac{1}{2}\left(\Delta_n(1-c')-|t|\log\left[\max\left(p^{2+3\delta}, nq\right)\right] - |t|(2+\delta)\log p\right)\right\}\\
&\quad\quad\preceq \text{exp}\left\{-\frac{1}{2}\left(\Delta_n(1-c')-w\gamma_n\right)\right\rbrace\to 0 \quad \text{as} \quad n \to \infty,
\end{align*}
where $w \in (0,1)$ and $c < 1-w$. It follows that
\[\sum_{k \in M_4}\text{BF}(k,t) \xrightarrow{\mathbb{P}}0.\]

\subsubsection*{Theorem 1 established}
We have thus shown that \[\sum_{k \in M_u}\text{BF}(k,t) \xrightarrow{\mathbb{P}}0,\]
for each of $u = 1,2,3,4$. \textcolor{black}{So we have $\mathbb{P}(\bm{Z} =t|\bm{Y}, \bm{X}, \sigma_\beta^2, \Sigma_Y) \xrightarrow{P} 1$ under the true data generating distribution for any $\sigma_\beta^2$ and $\Sigma_Y$, next we just need to integrate $\sigma_\beta^2$ and $\Sigma_Y$ out. Notice $q$ is some fixed integer, and $\sigma_\beta^2$ is independent with $\Sigma_Y$. Denote $\mathcal{D}$ as the collection of all $q\times q$ covariance matrices, $F_1(\cdot)$ as the distribution function of $\Sigma_Y$, $F_2(\cdot)$ as the distribution function of $\sigma_\beta^2$, then
\begin{equation*}
  \mathbb{P}(\bm{Z} =t|\bm{Y}, \bm{X}) = \int_{\mathbb{R}^+}\int_{\mathcal{D}} \mathbb{P}(\bm{Z} =t|\bm{Y}, \bm{X}, \sigma_\beta^2, \Sigma_Y) dF_1(\Sigma_Y)dF_2(\sigma_\beta^2).
\end{equation*}
Since $|\mathbb{P}(\bm{Z} =t|\bm{Y}, \bm{X}, \sigma_\beta^2, \Sigma_Y)|\leq 1$ and $\mathbb{P}(\bm{Z} =t|\bm{Y}, \bm{X}, \sigma_\beta^2, \Sigma_Y)\xrightarrow{P} 1$, by Dominated Convergence Theorem we have $\mathbb{P}(\bm{Z} =t|\bm{Y}, \bm{X}) \xrightarrow{P} 1$ \citep{shang2011consistency}. This establishes Theorem 1.
}
}





\section*{Declarations}

\subsection*{Data availability}
The authors confirm that all data underlying the findings are fully available without restriction. The phenotypic data and the SNP information of \emph{Oryza sativa} can be found at \url{https://www.nature.com/articles/ncomms1467}. The EFD coefficients of \emph{Oryza sativa} grain can be found at \url{https://www.ncbi.nlm.nih.gov/pmc/articles/PMC4380318/}.

\subsection*{Conflict of Interest}
The authors declare no competing interests.



\bibliographystyle{elsarticle-num}
\bibliography{references}

\end{document}